\documentclass[pdflatex,sn-basic]{sn-jnl}% Basic Springer Nature Reference Style/Chemistry Reference Style
\usepackage{graphicx}%
\usepackage{multirow}%
\usepackage{amsmath,amssymb,amsfonts}%
\usepackage{amsthm}%
\usepackage{mathrsfs}%
\usepackage[title]{appendix}%
\usepackage{xcolor}%
\usepackage{textcomp}%
\usepackage{manyfoot}%
\usepackage{booktabs}%
\usepackage{algorithm}%
\usepackage{algorithmicx}%
\usepackage{algpseudocode}%
\usepackage{listings}%
\makeatletter
\input{aas_macros.sty}
\def\ref@jnl#1{{\jnl@style#1\ }}
\makeatother

\newcommand{\dg}{^{\circ}}
\newcommand{\radm}{rad~m$^{-2}$}

\usepackage{comment}

\begin{document}

\title[The magnetic field of the Milky Way]{The magnetic field of the Milky Way: an observational perspective}


\author*{\fnm{Marijke} \sur{Haverkorn}}\email{m.haverkorn@astro.ru.nl}

\affil{\orgdiv{Department of Astrophysics/IMAPP}, \orgname{Radboud University}, \orgaddress{\street{PO Box 9010}, \postcode{6500 GL}, \city{Nijmegen}, \country{The Netherlands}}}

%ORCID: 0000-0002-5288-312X


%%% 150-250 words
\abstract{Magnetic fields are an important and enigmatic component of the Milky Way's ecosystem. Mostly frozen into interstellar plasma, they play key roles in (turbulent) gas dynamics, star formation, energy household, evolution of interstellar objects, and cosmic-ray propagation. This paper reviews recent progress on measuring and characterizing these Galactic magnetic fields, limited to the larger-scale fields in mostly diffuse media, and to an observational perspective.

On Galaxy-wide scales, the magnetic field roughly follows the spiral arms in the Galactic disk, and includes an additional component perpendicular to the disk away from the Galactic plane. The field configuration is different in the Galactic disk and the Galactic gaseous halo, qualitatively consistent with different dominating dynamo modes. Deviations from this idealized model are ubiquitously observed and include anomalously high Faraday rotation, variable magnetic field orientations and field reversals on kiloparsec scales.

On smaller scales, the magnetic field is turbulent, anisotropic and intermittent. Much used descriptions of the turbulent magnetic field such as power laws and Gaussianity are being replaced by higher-order statistics that better capture the complexities of the field. Magnetic field orientations and possibly strength are correlated with both cold and warm components of the multi-phase interstellar gas, and with the interstellar dust distribution.

The near future will bring a large increase in observational data in rotation measure grids, Faraday Tomography data and measurements of interstellar polarization of optical starlight, promising exciting developments in characterizing and understanding magnetic fields in the Milky Way in the next few years.}

%%%% STILL TO DO
%% All figures have descriptive captions (blind users could then use a text-to-speech software or a text-to-Braille hardware)
%% Patterns are used instead of or in addition to colors for conveying information (colorblind users would then be able to distinguish the visual elements)

% 4-6 keywords



\maketitle

\setcounter{tocdepth}{3} % TOC subsubsections
\tableofcontents

%%%%%%%%%%%%%%%%%%%%%%%%%%%%%%%%%%%%%%%%
\section{Introduction}

The Universe is rife with intriguing objects and processes, and astronomers and astrophysicists are unraveling their mysteries. Besides theory, numerical simulations and experiments, observations are the fourth indispensable key to understanding these mysteries - usually through electro-magnetic radiation, but increasingly also other messengers like particles or gravitational waves. One of the Universe's components that remains relatively hidden are magnetic fields, which do not emit observable radiation or particles at all.

Magnetism seemingly plays a very different role in space than on Earth. While daily life appears on the surface to be not much affected by magnetic fields, their effects are ubiquitous and abundant in space. The crucial difference is that most matter on Earth is electrically neutral, while practically the entire  Universe is ionized, from stars to cosmic filaments: even in the densest, darkest molecular cloud a small amount of free electrons remains. Although electric fields in space are quickly canceled out on larger scales due to the random configurations of positive and negative charges creating these electric fields, magnetic fields do maintain in the vast volumes of space, and can grow either extremely strong, or to extremely large spatial scales. Ionized gas directly interacts with these magnetic fields through the Lorentz force, effectively freezing in the field lines in plasma. Except for the densest molecular cloud cores with extremely low ionization degrees, all interstellar gas in galaxies and galaxy clusters has a sufficiently high ionization for the plasma to remain coupled to the magnetic field \citep{ferriere2001}. Because of this, magnetic pressure and/or magnetic tension play a major role in many astrophysical situations and the magnetic field is an active component in the dynamics and energy balance of many cosmic objects.

On galaxy-wide scales, magnetic pressure is a significant component in the Milky way's vertical hydrostatic equilibrium, its energy density comparable to cosmic ray and turbulent gas energy density \citep{boularescox1990%,fletchershukurov2001
}, although much smaller than galactic rotational energy \citep{heileshaverkorn2012}. Magnetic fields in the interstellar medium may influence the formation of galaxies \citep{naabostriker2017}, although likely only to a limited extent \citep{pakmoretal2017}. Galactic dynamics are affected by magnetic fields in e.g.\ the shaping of spiral arms \citep{gomezcox2002} and spurs \citep{kimostriker2002}, or the Milky Way's rotation curve at outer \citep{battaneretal1992%,ruizgranadosetal2012
} and inner radii \citep{chandelpopolo2022}.
Large-scale galactic magnetic fields influence the evolution of diffuse objects such as superbubbles \citep{%tomisaka1990,ferriereetal1991,
ntormousietal2017} or supernova remnants \citep{slavincox1992}%westetal2016}
, pulsar wind nebulae \citep{reynoldsetal2012}, planetary nebulae \citep{sabinetal2007}, and HII regions \citep{paveletal2012}. 
Magnetic fields obviously are an indelible part of magneto-hydrodynamic (MHD) turbulence, which is pervasive throughout the Universe. MHD turbulence is vital in many processes in the Milky Way, such as driving star formation \citep{maclowklessen2004}, or the maintenance of magnetic fields through dynamo action \citep{brandenburgntormousi2023}. Magnetic reconnection may be a significant heating source in the interstellar medium \citep{raymond1992}%,lazariancho2004}
.

% Beattie 2022: "Turbulence: However, the parameter space of MHD turbulence is large, and there need not be a universal phenomenology that captures the richness of the topic (see the eloquent review from Schekochihin 2020 about phenomenologies and two-point statistical models for incompressible MHD turbulence, which continue to be subject to debate; Iroshnikov 1964; Kraichnan 1965; Sridhar & Goldreich 1994; Goldreich & Sridhar 1995; Boldyrev 2006). "

%beck 2005:  Strong fields may shift the stellar mass spectrum towards the more massive stars (Mestel, 1990). --> 5 citations, none after 1994. Do not include.

%Pakmor et al 2017: the magnetic fields have only little effect on the global evolution of the galaxies as it takes too long to reach equipartition

In molecular clouds, magnetic pressure and tension significantly affect the gas dynamics at the scales probed by the Planck satellite \citep{planckintXXXV2016}. Cloud cores can collapse when they lose magnetic support, becoming magnetically supercritical, by processes like ambipolar diffusion of neutral and ionized species \citep{mestelspitzer1956,zweibel2015}, or diffusion of magnetic fields and plasma due to magnetic reconnection \citep{lazarian2014,santoslimaetal2021}. Increased magnetic tension in collapsing clouds counteracts this effect ('magnetic braking'), carrying angular momentum outwards which significantly slows down star formation \citep[e.g.,][]{mouschovias1976}.%,mestelparis1979,mouschoviaspaleologou1979,mouschoviaspaleologou1980}. 

In addition, magnetic fields dictate the propagation and spatial distribution of cosmic rays by deflecting their paths, and affect their spectra \citep{mollerachroulet2018}. Galactic (lower-energy) cosmic rays are completely randomized by the interstellar magnetic field \citep{strongetal2007}, while there is hope to trace back the paths of ultra-high energy cosmic rays (UHECRs) to reveal their origins, depending on their energy and charge \citep[e.g.][]{farrarsutherland2019,korochkinetal2025}.

%the acceleration and propagation of cosmic rays (Xu \& Yan 2013; Beattie et al. 2022a; Lazarian \& Xu 2022), 

As it is impossible to review the complete works on interstellar magnetic fields, and guided by my expertise and limitations therein, this review concentrates on insights on magnetic fields in the Milky Way on larger scales (largely ignoring magnetic fields in star formation and molecular clouds), and from an observational perspective. For readers who are interested in the aspects of magnetic fields not considered here, Sect.~\ref{s:reviews} presents an overview of other reviews on interstellar magnetic fields worth consulting. After an intermezzo in Sect.~\ref{s:terminology} explaining used terminology of magnetic field components, some background is provided on the origin, amplification, and maintenance of galactic magnetic fiels in Sect.~\ref{s:origin}. The most used observational tracers of magnetic fields are explained in Sect.~\ref{s:tracers} and Sect.~\ref{s:methods} treats the analysis methods using those tracers. Section~\ref{s:coherent} discusses magnetic fields on roughly kiloparsec scales, including models of the global structure of magnetic field in the disk and halo of the Milky Way, and observed large-scale and intermediate-scale deviations in those. Magnetic fields on smaller scales, turbulent fields and how magnetic fields interact with various other interstellar components, is described in Sect.~\ref{s:smallscale}. The magnetic field of the Local Bubble is arguably not a part of larger-scale Galactic magnetic fields; however, as it is relevant for many magnetic field measurements beyond the edge of the Local Bubble, a discussion on its magnetic field is included in Sect.~\ref{s:lb}. Finally, in Sect.~\ref{s:future}, I provide an overview of the major current and future efforts to observe magnetic fields, and close in the last section.

%%%%%%%%%%%%%%%%%%%%%%%%%%%%%%%%%%%%%%%%%%%%%%%%%%%%%%
\subsection{Reviews reviewed}
\label{s:reviews}

This review focuses on observational studies of interstellar magnetic fields, from approximately parsec scales to Galaxy scales, in typical diffuse interstellar environments, and focuses mostly on recent work. Therefore, many important studies of magnetic fields in the Milky Way will be missing. This section gives an incomplete but well-intended summary of where to find the information that is not discussed in this paper.

For mostly observational reviews on Galactic magnetic fields until about a decade ago, please see \citet{noutsos2013,haverkorn2015, han2017}. An overview of parametrized Galactic magnetic field models up to that time can be found in \citet{planckintXLII2016} and more recently in \citet{jaffe2019}. \citet{beckwielebinski2013}\footnote{Please refer to \url{https://arxiv.org/abs/1302.5663} for the latest updated version v11 in 2023.} review magnetic fields in the Milky Way and other galaxies. Extensive information on astrophysical magnetic fields is presented in the excellent book by \citet{shukurovsubramanian2021}. For Galactic magnetism as seen in the radio domain, see \citet{landecker2012} for an overview with emphasis on polarimetric surveys of diffuse Galactic emission, %detailing the information these surveys have yielded on small-scale magnetic fields, distances, and individual Galactic objects, 
and read \citet{akahorietal2018} for an outline of cosmic magnetism on scales from the interstellar medium to the cosmic web.

A theoretical review on galactic magnetic fields can be found in \citet{ferriere2015}, while \citet{brandenburgntormousi2023} provide a recent overview of galactic dynamo theory and simulations, and \citet{subramanian2019} discusses the origin and maintenance of coherent galactic magnetic fields, from primordial seed fields to galactic dynamos. 
Magnetic fields in and near the Galactic center are discussed in \citet{ferriere2011}, plasma turbulence and its tracers are explained in \citet{ferriere2020}, and magnetic fields in discrete Galactic objects, from radio loops and supernova remnants to HII regions and molecular clouds, are reviewed in \citet{han2017}. 
 For example \citet{vanlooetal2012, mckeeostriker2007,krumholzfederrath2019} discuss the role of magnetic fields in star formation. Magnetic fields in molecular clouds are reviewed from an observational perspective by \citet{crutcher2012} and by \citet{li2021}, while \citet{hennebelleinutsuka2019} focus on the numerical and theoretical aspects of magnetic fields in the formation and evolution of molecular clouds.
 
%Burkhart 2019: review "turbulence in neutral and molecular gas in galaxies"; has a Bfield section but not comprehensive enough to include. Too much off-topic.

%%%%%%%%%%%%%%%%%%%%%%%%%%%%%%%%%%%%%%%%
\section{Terminology of magnetic field components}
\label{s:terminology}

Galactic magnetic fields are exceedingly complex, and structured on a huge range of scales: from fields following spiral arms over many tens of kiloparsecs, to fields connected to interstellar turbulence on sub-parsec scales. On any range of scales, interstellar magnetic fields may be e.g.\ compressed by interstellar shocks, be (anti-)correlated with directions of filaments of ionized and/or neutral gas, be distorted by collapse of molecular clouds, or align with supernova remnant and superbubble walls. 

For characterization and description of these interstellar magnetic fields, they are traditionally divided into large-scale and small-scale components. The large-scale component, also often named coherent, regular, or uniform, is meant to denote the global field structure on scales of approximately kiloparsecs and higher. This includes spiral structure in the Galactic disk, and kiloparsec-scale fluctuations resulting from a Galactic dynamo in the halo. The small-scale (also called random, turbulent, incoherent) component describes the field on scales of roughly a (few) hundred parsecs and smaller, commonly described in terms of statistical parameters like power spectral index and correlation length.

\begin{figure}[ht]
    \centering
    \includegraphics[width=\textwidth]{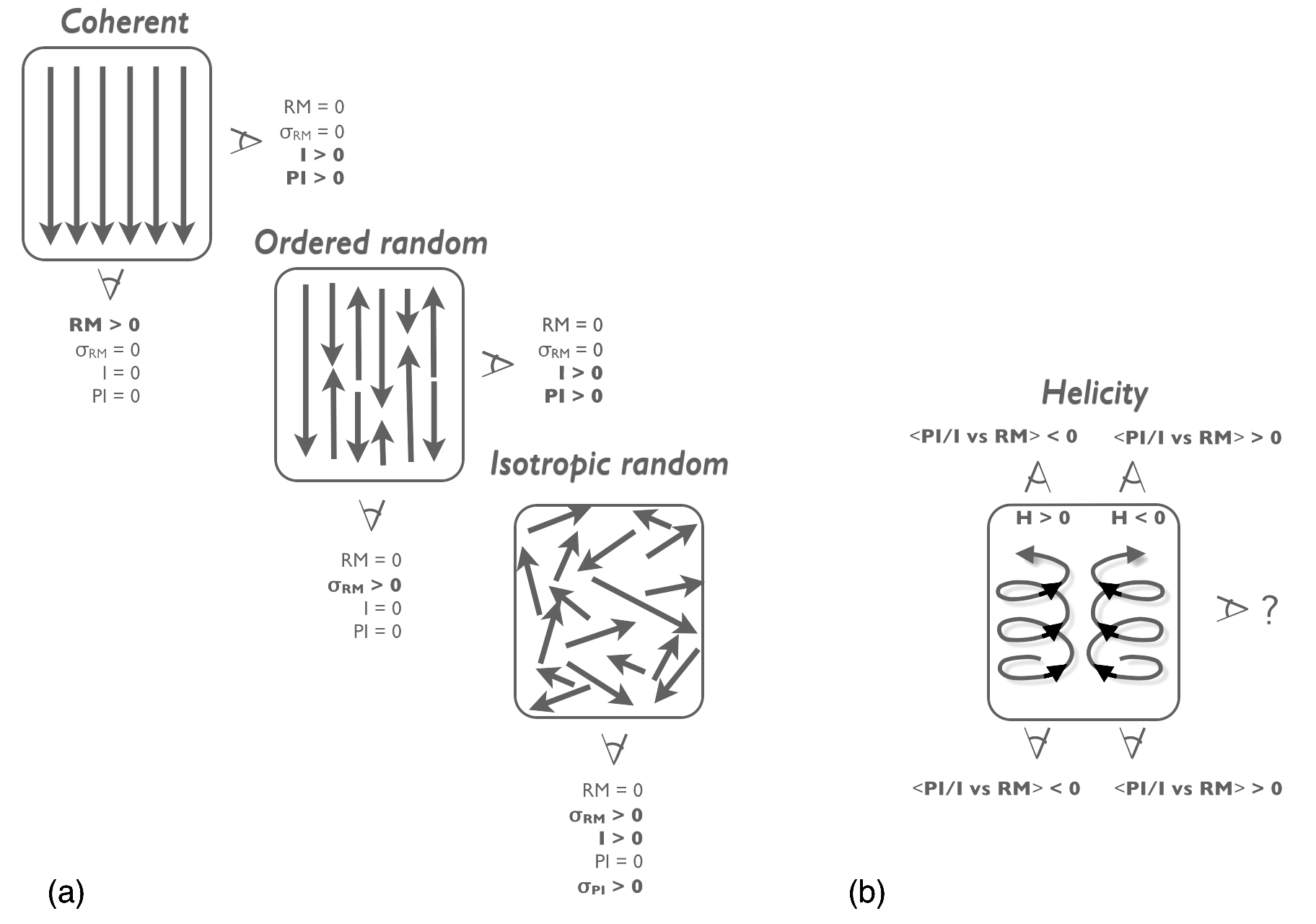} %%replaced with original from arXiv
    \caption{Overview of magnetic field components, and properties of radio-polarimetric observables. From left to right: sketch of a coherent field component, ordered random component and isotropic random component, and of a helical magnetic field. The eyes denote viewing directions, and some properties along this viewing direction of rotation measure (RM), dispersion of rotation measure ($\sigma_\mathrm{RM}$), total synchrotron intensity (I) and polarized intensity (PI). Reproduced from \citet{jaffe2019}, Galaxies, under a CC BY 4.0 license.}
    \label{f:fig1_jaffe19}
\end{figure}

Within the description of a turbulent magnetic field, an isotropic random (also called ordered random, or striated) field and an anisotropic random component can be distinguished (see Fig.~\ref{f:fig1_jaffe19} and the extensive explanation in \citealt{jaffe2019}). Anisotropic random magnetic fields contain small-scale structure in direction, but not in orientation. These fields can arise from isotropic random magnetic field structure as a result of shear due to Galactic differential rotation, density waves in spiral arms, or compression e.g.\ due to the passage of a shock wave. This component can be determined by the different response of isotropic and anisotropic turbulence to measured total intensity, polarized intensity and rotation measure, see Fig.~\ref{f:fig1_jaffe19}.

This distinction is useful, although it ignores many complexities. First of all, meso-scale structure is hard to classify, such as kiloparsec-scale deviations from spiral arm structures. Also, magnetic helicity, a measure of the twisting and curling of the field, is missing in this simple description (see Fig.~\ref{f:fig1_jaffe19}). Lastly, correlations of magnetic field with discrete interstellar gas or dust structures such as interstellar shocks, clouds or filaments are usually not adequately described by this distinction, especially when they are phenomenologically studied on a case by case basis. 

This paper contains a slightly more detailed distinction. Coherent, large-scale fields are discussed separately for the Galactic disk (Sect.~\ref{s:coherent-disk}) and the Galactic halo\footnote{The 'Galactic halo' in the interstellar medium and Galactic magnetism communities is differently defined as (dark matter dominated) galaxy haloes in galaxy formation. Here, the Galactic halo is described as the layers of interstellar medium above the thin Galactic star-forming disk, containing the hot coronal gas, the Reynolds' Layer of ionized gas, neutral gas and dust, magnetic fields and cosmic rays beyond the Galactic disk.} (Sect.~\ref{s:coherent-halo}). In the discussion of small-scale magnetic fields (Sect.~\ref{s:smallscale}), distinction is made between a description in terms of turbulence in Sect.~\ref{s:ss-turb} and correlations with interstellar gas and dust structures in Sect.~\ref{s:ss-comp}.

%%%%%%%%%%%%%%%%%%%%%%%%%%%%%%%%%%%%%%%%
\section{Origin, amplification and maintenance of galactic magnetic fields}
\label{s:origin}

In galaxies (and most other astrophysical contexts), the electrical conductivity of the plasma is so high (i.e.\ magnetic Reynolds number is so large) that Ohmic resistivity can be neglected at most scales. This would mean that magnetic fields present in this plasma would survive indefinitely. However, due to turbulent plasma motions, the magnetic field fluctuations are rapidly transported to small scales, where magnetic and kinetic energy is converted into heat, producing decay of the magnetic field. However, magnetic fields on the order of a few microGauss  at galactic scales are known to exist already at redshift $\sim1.3$ \citep{bernetetal2008}, which means that a mechanism must exist that maintains these magnetic fields despite the fact that galaxies and protogalaxies are turbulent. This mechanism is dynamo action, which converts kinetic (and to a smaller scale gravitational) energy into magnetic energy. Dynamos can amplify small magnetic fields to microGauss strength, but cannot create magnetic fields in a non-magnetized plasma. Therefore, some kind of weak magnetic seed field is needed for the dynamo to act.

Dynamos can be classified into two categories: mean-field dynamos and fluctuation dynamos (also commonly called large-scale and small-scale dynamos). Mean-field dynamos can amplify and maintain magnetic fields on spatial scales larger than the correlation length of the turbulent flows involved, while the fluctuation dynamo only amplifies fields on scales smaller than that. 
In the early phases of a dynamo, magnetic fields are still so weak that their feedback is negligible: the induction equation is linear in the magnetic field, which means that field grows exponentially; this is the linear or kinematic regime. When the field strength increases, feedback on the dynamo process becomes significant and the dynamo enters a non-linear regime. As the time scales for magnetic field growth in spiral galaxies are very small for fluctuation dynamos ($\lesssim 10$~Myr), but also not long in mean-field dynamos ($\sim 0.5-1$~Gyr), we expect that the dynamos acting in the Milky Way and other low-redshift spiral galaxies are in the non-linear, saturated regime \citep[][hereafter SS21]{shukurovsubramanian2021}. %However, if the non-linear effects are weak, it is possible that some effects of seed fields are still preserved, primarily in the outer regions of the Galaxy CHECK. E.g. reversals say SS21, work out.

%%%%%%%%%%%%%%%%%%%%%%%%%%%%%%%%%%%%%%%%
\subsection{Mean field dynamos}
%mean-field dynamos
%Quadrupolar modes grow faster than dipoler in galaxies (sectio 11.3 SS21)
The mean field dynamo process was first described by \citet{parker1955} and first applied to a disk galaxy by \citet{parker1971} and \citet{vainshteinruzmaikin1971}. Assuming (near or perfect) flux freezing, differential rotation of the magnetized interstellar plasma will shear any poloidal magnetic field component into a toroidal component (the $\Omega$-effect). Turbulent motions in a stratified plasma combined with the Coriolis force can convert this toroidal component back in a poloidal component (the $\alpha$-effect). These turbulent motions are thought to be mostly driven by supernova explosions in the disks of galaxies (SS21), possibly enhanced by the magneto-rotational instability \citep{balbushawley1991} in the outer parts where supernova driving is weaker or absent \citep{piontekostriker2007}. Magnetic buoyancy, amplified by cosmic rays, enhances the creation of poloidal motions \citep{parker1992,qazietal2025}.
Combined, the differential rotation and helical turbulence can amplify galactic magnetic fields in a so-called $\alpha-\Omega$ dynamo mechanism to the observed microGauss strengths. 

In the Milky Way, there is ample evidence that a mean-field dynamo is at work. The observed magnitude and sign of the pitch angle of the magnetic spiral arms, the quadrupolar ('butterfly pattern') signature of magnetic field direction in the Galactic halo, and large-scale reversals in the direction of the disk magnetic field (see Sect.~\ref{s:coherent}) are all consistent with a mean-field dynamo in the Milky Way \citep{shukurov2004}. 
%Dynamos can also be driven by magnetic buoyancy. Magnetic field in the disk is buoyant as pressure balance dictates that regions of higher magnetic field have lower gas density. Cosmic rays can also increase magnetic buoyancy (parker hanasz). vertical buoyant motions become helical due to coriolis force
%twisting of primordial field by diff rotation would give much smaller pitch angles than dynamo and thean observed. quadrupolar symmetry is consistent with dynamo theory. magnetic arms in material interarm regions indicate deviations from flux freezing and thus for dynamo action REWORD as is from SS21. because if not frozen in then will decay and need dynamo

% feedback effects can be alpha-quenching, where the alpha effect is assumed to be a function of magnetic field strength, giving a decrease in alpha effect with growing magnetic field strength; which gives a magnetic field growth rate with a certain dependence on magnetic field strength.

%also then need to remove magnetic helicity from the system (see 7.12 SS21), e.g. with galactic outflows (shukurov et al 2006; sur et al 2007a, chamandy et al 2014b) or turbulent diffusion (kleeorin et al 2000, 2003, 2003, 2006)

%%%%%%%%%%%%%%%%%%%%%%%%%%%%%%%%%%%%%%%%
\subsection{Fluctuation dynamos}

A fluctuation dynamo creates small-scale, random magnetic fields, from random plasma motions in a medium with high magnetic Reynolds number such as the interstellar medium  \citep{kazantsev1968}. It can amplify magnetic field fluctuations exponentially, through stretching and folding magnetic field lines in interstellar turbulence on increasingly smaller scales. The produced magnetic fields are intermittent and locally concentrated into intense thin filaments (of a length comparable to the turbulent scale) surrounded by a less ordered magnetic field. A fluctuation dynamo is thought to maintain magnetic fields in astrophysical situations where the mean-field dynamo is absent due to lack of differential rotation (e.g.\ in galaxy clusters), but is also active in the Milky Way's interstellar medium, possibly as a seed mechanism for the mean-field dynamo (SS21).
%Time scale of magnetic field growth in fluctuation dynamo is $\lesssim 10^7$ years in spiral galaxies (SS21, p191) in exponential growth (kinematic dynamo) so hard to observe. Only saturated state of fluctuation dyamo observable.

%%%%%%%%%%%%%%%%%%%%%%%%%%%%%%%%%%%%%%%%
\subsection{Seed magnetic fields}
A dynamo can amplify magnetic fields, but not create fields in a non-magnetized plasma. Therefore, a minute seed field needs to be generated by other processes, for a dynamo to amplify. 
In primordial magnetogenesis, magnetic fields can be created through a number of exotic physical processes such as quantum fluctuations of the electro-magnetic field during inflation or in phase transitions. 

In the inflation era, electromagnetic quantum fluctuations could be transformed to classical fluctuations and stretched to large scales. However, finetuning of parameters seems to be necessary, and it is as yet unclear how to amplify electromagnetic wave fluctuations to a strong enough seed magnetic field, which requires breaking conformal invariance of the electromagnetic field \citep{turnerwidrow1988,subramanian2016}.
Primordial magnetic fields may also be created in various phase transitions, such as the electroweak  or quark-hadron phase transitions \citep{hogan1983,vachaspati2021}. This explanation is appealing if these processes create fields on large spatial scales in the present, but it is not evident that strong enough seed fields can be created to enable significant amplification of magnetic fields on sufficiently small timescales \citep{kandusetal2011}.

More traditional physics can create small magnetic fields through some battery mechanism, in which positive and negative charges are separated, creating an electric field, which under certain circumstances can create a magnetic field. The most well-known of these mechanisms is the Biermann battery \citep{biermann1950}. Here, a pressure gradient in the plasma accelerates the electrons faster than the far more massive protons, creating an electric field. If there is a density gradient that is not aligned with the pressure gradient, this generates a weak magnetic field. Battery mechanisms can only grow magnetic fields linearly with time, and can only produce very small seed fields.

Plasma instabilities in the formation of galaxies and galaxy clusters from cosmological density perturbations can also provide a seed field. An example is the Weibel instability \citep{weibel1959}, which is created in anisotropic velocity distributions of the plasma. Even though the Weibel instability amplifies magnetic fields exponentially, the fields are on such small scales that this process is not believed to create a relevant seed field at galactic scales. The same objections hold for seed fields created in smaller objects such as active galactic nuclei or even stars and then ejected into the interstellar gas, which can amplify magnetic fields fast but on too small scales. A fluctuation dynamo is a viable option to amplify minute seed fields to the strengths that a mean-field dynamo can amplify to and maintain at the observed microGauss strengths of galactic magnetic fields. For details, see SS21.
%turner widrow 1988 seminal paper on field creation in inflation era

%%%%%%%%%%%%%%%%%%%%%%%%%%%%%%%%%%%%%%%%
\section{Observational tracers}
\label{s:tracers}

%%%%%%%%%%%%%%%%%%%%%%%%%%%%%%%%%%%%%%%%
\subsection{Polarized dust emission and absorption}

Both, \citet{hall1949} and \citet{hiltner1949} semi-independently discovered that starlight can be partially linearly polarized, and suggested from the uniformity of polarization directions in nearby stars that this polarization was not intrinsic to the star but caused by the intervening interstellar medium.

\begin{figure}[ht]
    \centering
    \includegraphics[width=0.8\textwidth]{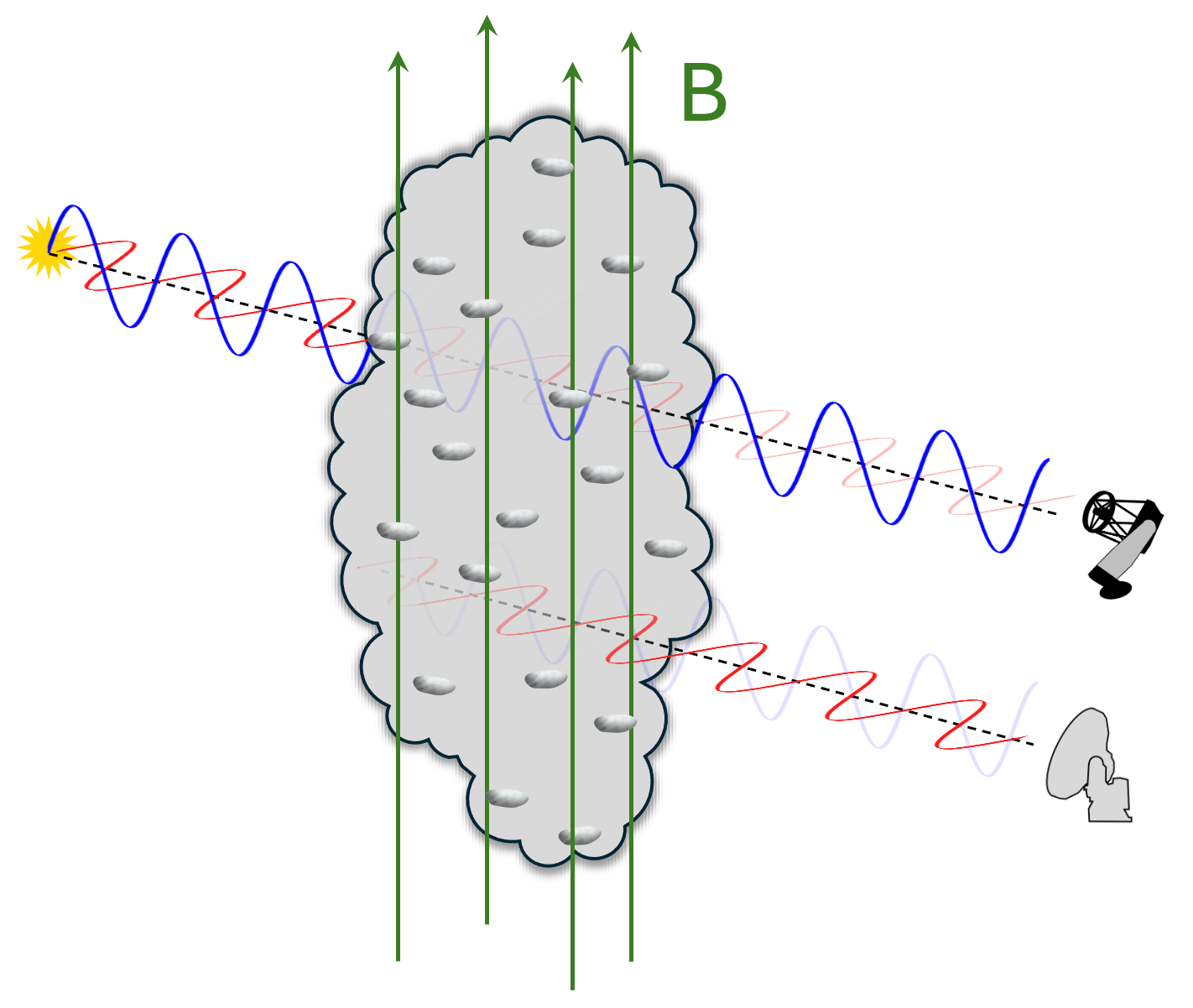}
\caption{Effects of polarized radiation by interstellar dust grains in an ambient magnetic field. Top: unpolarized light from a background star attains partial (optical and NIR) linear polarization by propagating through a magnetized dusty medium, which is parallel to the magnetic field orientation projected on the plane of the sky. Bottom: dust grains attain (FIR and submm) partial linear polarization in an ambient magnetic field, which is perpendicular to the magnetic field orientation projected on the plane of the sky.}
    \label{f:starlightpol}
\end{figure}

Asymmetric dust grains with some angular momentum in an ambient magnetic field will align their angular momentum with the local magnetic field orientation, see Figure~\ref{f:starlightpol}.  Dichroic absorption of starlight propagating through this medium will then incite partial linear polarization \citep{stein1966}, which is directed parallel to the plane-of-sky magnetic field component averaged along the dust-weighted line of sight. In the following, we refer to this as interstellar polarization of starlight. The alignment efficiency can be low in very dense media, but can in most interstellar gas be assumed to be 100\%: the relatively high ratio of polarization over extinction \citep{jones1996}, the wavelength dependence of polarization \citep{mathis1986} and more complex dust population models \citep{kimmartin1995} are all best explained with high alignment efficiency.

The mechanism causing alignment was long thought to be due to the Davis--Greenstein effect \citep{davisgreenstein1951}, but is currently best explained by the Radiative Alignment Torque (RAT) theory, giving torques due to asymmetries in the grain shapes and in ambient radiation fields \citep{purcell1975, lazarianhoang2007}, possibly in combination with the Davis--Greenstein effect. Note however that in exceptional cases, alignment mechanisms may cause alignment perpendicular to the local plane-of-sky field orientation \citep{raoetal1998}.

Aligned dust grains will not only incite a partial linear polarization in (optical and infrared) passing starlight, but also emit thermal far infrared (FIR)/submillimeter (submm) emission which is partially linearly polarized, oriented \emph{perpendicular} to the plane-of-sky magnetic field component averaged along the dust-weighted line of sight, i.e.\ perpendicular to the orientation of interstellar polarization of starlight by the same medium, as demonstrated in Fig.~\ref{f:starlightpol}. 

The amount of polarization carries information on the orientation of magnetic field in dusty environments and of the dust properties, while the polarization angle traces the line-of-sight integrated and dust-weighted magnetic field orientation projected on the plane of the sky. The two observational methods mentioned here are nicely complementary: dense environments are typically observable in submm and FIR polarized thermal dust emission, due to the high amounts of dust along a line of sight, where extinction is so high that polarized background stars are scarce. In more diffuse environments, where dust emission is weak, optical and near infrared (NIR) polarization in background stars is measurable.

%--------------------------------------
%--------------------------------------

%%%%%%%%%%%%%%%%%%%%%%%%%%%%%%%%%%%%%%%%
\subsection{Synchrotron emission}

Relativistic cosmic-ray electrons in galaxies circling around magnetic field lines under influence of the Lorentz force emit synchrotron emission. A power law distribution of energies of the relativistic electrons translates into a power law in emission. At the low field strengths and ubiquitous cosmic-ray presence in the interstellar medium, this emission is pervasive and is detected in the submm to radio domain \citep{fermi1949}. 

Synchrotron emission is intrinsically highly linearly polarized in the direction of the local magnetic field component perpendicular to the line of sight, but is partially depolarized by the integration along the line of sight and within the observing beam in case of a non-uniform magnetic field orientation. At high radio (microwave) wavelengths, where Faraday rotation is negligible (see Sect.~\ref{s:frot}), the  observed orientation of linear polarization angle represents the orientation of the plane-of-sky magnetic field component averaged over the path length. The degree of polarization gives information on the tangling of the magnetic field along the line of sight \citep{beckwielebinski2013}. The synchrotron intensity can also be a measure of the magnetic field strength under certain energy assumptions such as local equipartition or pressure equilibrium, although these assumptions may be problematic \citep{setabeck2019}. Measuring total and polarized synchrotron emission is an excellent method to estimate magnetic field strengths and the amount of tangledness of the field in external galaxies, and is one of the main tracers used in large-scale modeling of the Milky Way's magnetic field (see Sect.~\ref{s:coherent}). Synchrotron emission is also used to trace the properties of the magnetized turbulent magnetic field component through various analysis methods, see the review by \citet{zhangwang2022}.

%%%%%%%%%%%%%%%%%%%%%%%%%%%%%%%%%%%%%%%%
\subsection{Faraday rotation}
\label{s:frot}

Faraday rotation is the rotation of linear polarization angle in a magnetized and ionized medium, caused by different phase velocities for left and right hand circular polarized radiation in magnetized plasma, called circular birefringence. The Faraday rotation of a linearly polarized wave rotates its polarization angle $\chi$ in an intervening medium with thermal electron density $n_e$ and magnetic field $\mathbf{B}$, by an amount:
\[
\Delta\chi = \chi_0 + \mbox{RM}\lambda^2 = \chi_0 + \left[0.81\int_\mathrm{source}^0 \left(\frac{n_e}{\mbox{cm}^{-3}}\right)\left(\frac{\mathbf{B}}{\mu\mbox{G}}\right)\cdot \left(\frac{\mathbf{ds}}{\mbox{pc}}\right)\right]\lambda^2,
\]
where $\chi_0$ is the intrinsic polarization angle at emission, $\lambda$ the observing wavelength, and RM the rotation measure. For emission sources behind the Faraday rotating medium, RM can be easily determined as the linear relation between $\Delta\chi$ and $\lambda^2$. However, when synchrotron emitting and Faraday rotating media are mixed, RM is not defined anymore. Instead, the medium can be described by the Faraday spectrum, which depicts polarized intensity as a function of Faraday depth defined as 
\[
\left(\frac{\phi(d)}{\mbox{rad~m}^{-2}}\right) = 0.81\int_{d}^0 \left(\frac{n_e}{\mbox{cm}^{-3}}\right)\left(\frac{\mathbf{B}}{\mu\mbox{G}}\right)\cdot \left(\frac{\mathbf{ds}}{\mbox{pc}}\right),
\]
where $d$ is the distance to a particular component of the synchrotron emission. 

If estimates of the mean free electron density and of the path length of the intervening ionized medium can be independently obtained, Faraday rotation provides an estimate of the strength of the magnetic field component directed along the line of sight, either path length integrated (RM) or in a spectrum ($\phi(d)$). This calculation assumes that electron density and magnetic field strength are uncorrelated. If pressure equilibrium holds, magnetic field and electron density will be anticorrelated, in which case the magnetic field strength will be underestimated; on the other hand, a positive correlation will overestimate magnetic field strength \citep{becketal2003}. Even if reliable estimates of free electron density and/or path length are not available, the sign of the RM still carries direct information on the direction of the (path length integrated) line-of-sight component of the magnetic field. Analysis methods using Faraday rotation will be discussed in Sections~\ref{s:rmgrids} and \ref{s:fartom}.

%%%%%%%%%%%%%%%%%%%%%%%%%%%%%%%%%%%%%%%%
\subsection{Zeeman effect}

The Zeeman effect splits a spectral line into several components of slightly different frequency under influence of a magnetic field \citep{zeeman1897}. The transversal Zeeman effect creates two linearly polarized lines proportional to the plane-of-sky magnetic field component; this is regularly observed in the Sun \citep{hale1908}.
 The longitudinal Zeeman effect splits the line into two circularly polarized components proportional to the magnetic field component parallel to the line of sight. As opposed to the transversal Zeeman effect, the longitudinal effect can be detected in the interstellar environment through its circular polarization signal, however still  only in the coldest gas. Zeeman splitting is therefore complementary to many other methods, in that it gives an \textit{in situ} measurement of magnetic field component along the line of sight in the cold dense interstellar medium. It is very challenging, as the high signal-to-noise ratio required for this weak effect necessitates long observing times. However, it is an important method to measure magnetic field strengths in molecular clouds \citep[e.g.,][]{crutcheretal2010}.

%%%%%%%%%%%%%%%%%%%%%%%%%%%%%%%%%%%%%%%%
\subsection{Other tracers} 

The \emph{Goldreich--Kylafis effect} \citep{goldreichkylafis1982} is a quantum mechanical effect that predicts linearly polarized spectral lines from rotational transitions under influence of a magnetic field parallel to the line of sight. Polarization would only arise if magnetic sublevels of rotational level $J=1$ are unequally populated e.g.\ due to an anisotropic radiation field \citep{goldreichkylafis1981}. In addition, the optical depth of the cloud should be around unity, and radiative transition rates should be comparable to or dominate over collisional rates -- conditions that are not uncommon in the interstellar medium. The Goldreich-Kylafis effect was first detected in a molecular outflow from a protostar \citep{girartetal1999} and later found in a variety of molecular sources.
%e Lai et al. (2003), Girart et al. (2004), Cortes et al. (2005), Forbrich et al. (2008), Cortés et al. (2021), Barnes et al. (2023), from Tritsis & Kylafis 2025

% observations of Goldreich-kylafis: https://ui.adsabs.harvard.edu/abs/2008A%26A...492..757F/abstract

A similar process is \emph{Ground State Alignment (GSA)}, the alignment of angular momentum of atoms and ions with fine or hyperfine structure by external radiation fields and realignment through precession in a magnetic field \citep{yanlazarian2006}. %, yanlazarian2007,yanlazarian2012}. 
This atomic alignment induces a few percent polarization in interstellar absorption lines of e.g.\ O~{\sc i}, S~{\sc ii} or Ti~{\sc ii}. %and is the only case of linear polarization in UV lines. 
It is sensitive to very weak fields in dilute media, and uniquely informs about the local 3D magnetic field. It has been successfully used in stellar systems \citep{zhangetal2020} and supernovae \citep{yangetal2022}%, vasylyevetal2024}
, but is not yet observed in the interstellar medium.

\citet{houdeetal2009_molionspectra} reviews the determination of magnetic field strength based on \emph{line width variations of co-spatial ionic and neutral molecules}. Molecular lines from ionized species are observed to be narrower than lines from co-spatial neutral species \citep{houdeetal2000}. This is thought to be due to ambipolar diffusion in the turbulent, magnetized multi-phase medium, allowing determination of the local magnetic field strengths \citep{lihoude2008}. 

% see recent polstar paper (Andersson et al 2022) confirmed no IS detections.

%Beck 2010: predicts a polarization e.g. in the optical and UV lines of Oi, Sii and Tiii, but has not yet been detected
%" linear polarization of interstellar absorption or fluorescence emission lines from ions pumped by an anisotropic illuminating flux."
%"The atoms get aligned in terms of their angular momentum and, as the life-time of the atoms/ions we deal with is long, the alignment induced by anisotropic radiation is susceptible to very weak magnetic fields (1G > B > 10−15G, Yan & Lazarian 2012), which is exactly the level of the magnetism in diffuse medium, including both ISM and IGM."

%%%%%%%%%%%%%%%%%%%%%%%%%%%%%%%%%%%%%%%%
\section{Analysis methods}
\label{s:methods}

Since magnetic fields can solely be observed indirectly, a multitude of methods exist to interpret the various polarization measurements discussed above, in terms of magnetic fields. This Section summarizes the most used of these.

% OTHER METHODS:
%Yue Hu (2025 ApJ 990, 76): ML to get 3D magnetic field from HI cubes, using MHD simulations as training. Uses CRAFTS observations in Monoceros region with an IVC and LVC cloud.

%%%%%%%%%%%%%%%%%%%%%%%%%%%%%%%%%%%%%%%%
\subsection{Davis--Chandrasekhar--Fermi method}

\citet{davis1951} and \citet{chandrasekharfermi1953} realized that one could estimate the magnetic field strength in gaseous media through the dispersion in observed polarization angle, under certain assumptions about the medium and the field. Assuming isotropic, incompressible Alfv\'enic magnetic field fluctuations in the plane of the sky, and assuming small fluctuations, the dispersion in observed polarization angle in submm dust emission or optical/NIR dust absorption measurements is then related to the magnetic field strength in the plane of the sky as 
\[
\sigma_\theta = \frac{\sigma_b}{B_0},
\]
with $\sigma_b$ the dispersion in turbulent magnetic field strength $b$, and $B_0$ the regular magnetic field strength. Assuming that turbulent magnetic energy density and turbulent kinetic energy density are in equipartition, the turbulent magnetic field strength can be described in terms of the gas density $\rho$ and velocity dispersion $\sigma_v$ as:
\[
\frac{1}{2}\rho \sigma_v^2 = \frac{\sigma_b^2}{8\pi},
\]
Combining these expressions leads to the Davis--Chandrasekhar--Fermi (DCF) equation
\[
B_0 = \sqrt{4 \pi \rho}\frac{\sigma_v}{\sigma_\theta}.
\]

%Myers & Goodman (1991) and Houde et al. (2009) have pointed
%out that if the turbulent correlation length, δ, is less than the thickness
%of the region being observed along the LOS, w, then the dispersion
%in PAs will be reduced.

The DCF mechanism is a quite tricky method to estimate the interstellar magnetic field, as the necessary assumptions are stringent and often not applicable in the interstellar medium. Many efforts have been made to adapt the original DCF formalism to relieve these assumptions and allow for less stringent additions. The assumptions, followed by mitigation efforts, are listed below:
\begin{itemize}
\item perturbations in the magnetic field are much smaller than the mean field, so that the small-angle approximation holds \citep{ostrikeretal2001}. Accuracy increases by replacing the dispersion in polarization angle $\theta$ by the dispersion of $\tan\theta$, $\sigma(\tan\theta)$, \citep{heitschetal2001} or using the tangent of the angle dispersion, $\tan(\sigma_\theta)$  \citep{falcetaetal2008,lietal2022}. Alternatively, one can express the angle dispersion in terms of the Stokes parameters $Q$ and $U$ to avoid the small-angle approximation \citep{zweibel1996}. 
\item the velocity fluctuations are all due to turbulence. However, in reality there will be contributions from the larger-scale magnetic field to the angular dispersion, and non-turbulence contributions to linewidths. Larger-scale fluctuations in the regular magnetic field can be simple modelled and subtracted \citep{girartetal2006}, or removed by spatial filtering of the polarization angle map \citep{pattleetal2017, pillaietal2015}. Undesired contributions  to the linewidth from thermal broadening can be avoided by using the positional fluctuations in the velocity centroids of a spectral line, i.e.\ the first moments of each spectrum \citep{esquivellazarian2005,kandeletal2017}, since thermal broadening is symmetric around a line center and therefore does not affect velocity centroids. 

\item turbulent magnetic field fluctuations are isotropic and Alfv\'enic, i.e.\ cause transverse incompressible waves in the magnetic field lines. However, the interstellar magnetized gas is highly compressible. %Neglecting compressible waves overestimates the turbulent velocities, and therefore overestimates magnetic field strengths. 
\citet{skalidistassis2021} describe an attempt to include compressible waves; also see \citet{lazarianetal2022} for a discussion about their assumptions.
\item equipartition between kinetic and magnetic turbulent energies holds. If actual equipartition does not hold, or includes also other energetic components such as thermal or cosmic ray pressure or external gravity sources, the magnetic field will be overestimated \citep{zweibel1996}.
\item the maximum angular scales of turbulent velocity and magnetic field fluctuations are smaller than the field of view. If this is not the case, the dispersion in velocity and polarization angle are only lower limits to the true dispersion. In this case, one can use the structure function (SF, \citet{falcetaetal2008}) or the angular dispersion function (ADF, \citet{hildebrandetal2009, houdeetal2009})\footnote{The ADF is defined as the square root of the structure function, so that this method is also sometimes referred to as the Structure Function DCF (SF-DCF or DCF/SF) method.} to estimate angle fluctuations on a range of scales.  Structure functions of both dust emission polarization angle as well as of turbulent velocities can be compared to MHD simulations, to obtain estimates of small-scale magnetic fields in an approach called Differential Measure Analysis (DMA, \citet{lazarianetal2022}).
%A special case of this is treated in \citet{hildebrandetal2009)}, who derive an expression for the turbulent to regular magnetic field ratios for situations where the maximum scale of turbulence is smaller than the ADF resolution.
\item the dusty medium is homogeneous. If the medium is clumpy, magnetic field estimates will deviate, with stronger magnetic fields inferred for large covering factors of the clumps \citep{zweibel1990}.
\item no ambipolar diffusion. The effect of ambipolar diffusion will decrease the tangling of the magnetic field, thus causing an overestimate of the DCF magnetic field strength \citep{zweibel1990}.
\item the path length through the dusty medium is much shorter than the line of sight, as integration along the line of sight can decrease the polarization angle dispersion by depolarization \citep{myersgoodman1991}.  \citet{choyoo2016} provide a correction to the DCF equation, viz the square root of the ratio of turbulence driving scale over line of sight length, to take into account line of sight averaging.
\end{itemize}

In addition, observational constraints will alter DCF results: beam-smoothing of polarized submm dust emission will decrease the angle dispersion, causing overestimations of magnetic field strengths \citep{heitschetal2001,falcetaetal2008, houdeetal2009}, and interferometric filtering of large spatial scales causes underestimation of magnetic field strengths, which however can be modeled and accounted for \citep{houdeetal2016}.

Because of the large number of assumptions and simplifications inherent to the DCF methods, tesing DCF with numerical simulations is a crucial endeavor. A first-order correction to the DCF-obtained magnetic field strengths is introducing a fudge factor $f$ that absorbs errors due to effects like medium-scale inhomogeneities, anisotropies, and averaging along the line-of-sight, as:
\[
B_0 = f \sqrt{4 \pi \rho}\frac{\sigma_v}{\sigma_\theta}.
\]
Various MHD simulations for cold compressible medium in giant molecular clouds \citep{ostrikeretal2001,heitschetal2001,padoanetal2001,kudohbasu2003,liuetal2021} determine suitable fudge factors $f$ generally between $\sim0.4$ and $\sim0.8$. Generally, the DCF-approximation works better for strong fields than for weak fields \citep{heitschetal2001,liuetal2021}. 
 
%Liu et al 2021:  We test the Cho \& Yoo (2016) method and find that this method correctly accounts for the effect of line-of-sight signal integration at $>0.1$ pc, but fails at $<0.1$
%The fudge factor depends on resolution (Heitsch).
%Padoan et al. 2001: test simulations, obtain fudge factor f ~ 0.4

%\mh{Zweibel (1990), Meyers \& Goodman (1991):
%\[
%B = \xi \sqrt{\frac{4}{3}\pi\rho}\frac{v_{rms}}{\Delta\psi},
%\]
%where $\xi$ is a fudge factor taking into account \mh{[quote F-G08]} "medium inhomogeneities, anisotropies on velocity perturbations, observational resolution and differential averaging along the LOS."}

%%%%%%%%%%%%%%%%%%%%%%%%%%%%%%%%%%%%%%%%
\subsection{Rotation Measure (RM) grids}
\label{s:rmgrids}

Faraday rotation measurement of background sources is an extremely versatile way to probe magnetic fields in the gaseous medium in between the source and observer. Linearly polarized extragalactic background sources such as quasars and radio galaxies mostly occur as point sources on the sky at the relatively low spatial resolutions of large-coverage radio surveys. Their measured rotation measure is a combination of any contributions from the source itself (intrinsic), from the circumsource medium, from the intragalactic medium (IGM), from the Galaxy, and from the Earth's ionosphere (plus intracluster medium if appropriate):
\[
RM_\mathrm{meas} = RM_\mathrm{intr} + RM_\mathrm{cs} + R_\mathrm{IGM} + R_\mathrm{MW} + RM_\mathrm{ion} (+ RM_\mathrm{intracl}).
\]
Large angular-scale correlations imply local effects, generally dominated by the Milky Way. 
At all but the highest Galactic longitudes, the contribution from magnetic fields in the Milky Way dominates \citep{simardnormandinkronberg1980}. RMs from intergalactic filaments, galaxy groups, clusters and superclusters are measured to be of the order of 1~to a few~\radm\ \citep[e.g.][]{johnstonhollitekers2004,%xuetal2006,akahoriryu2010,vaccaetal2018,andersonetal2024,
carrettietal2025}, and RMs intrinsic to the extragalactic radio source are estimated to be $\lesssim 7$~\radm\ \citep{oppermannetal2015}. 

\begin{figure}[ht]
    \centering
    \includegraphics[width=\textwidth]{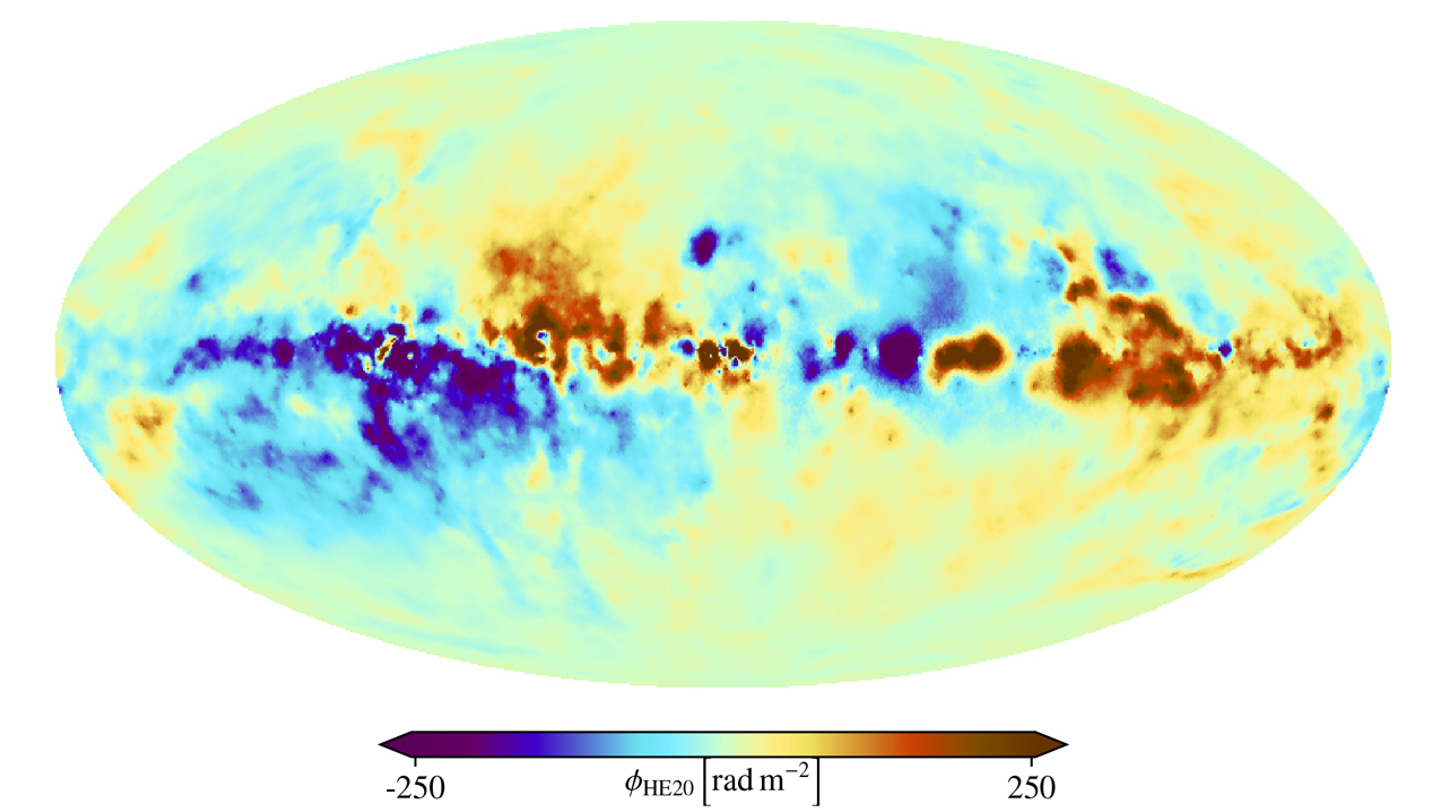}
    \caption{Galactic Faraday rotation measures over the whole sky, derived from extragalactic source RMs where intrinsic source contributions have been removed. Figure reproduced from \citet{hutschenreuteretal2022} (CC BY 4.0).}
    \label{f:hutsch22}
\end{figure}

The ionospheric contribution to the RM is up to a few \radm\ depending on time of day and Solar activity, and can be calibrated out in data processing.  RM contributions from the magneto-ionized medium intrinsic to the extragalactic sources and/or their surroundings are expected to be independent from source to source and therefore can be corrected for or removed. The simplest method to do this is simply (weighted) averaging over a large number of sources \citep{tayloretal2009,pshirkovetal2011,xuhan2014RAA}, but more sophisticated methods are used like spherical harmonics decomposition \citep{dineencoles2005}, or various methods taking into account uncertainties in the noise covariance \citep{shortetal2007,oppermannetal2012,oppermannetal2015,hutschenreuterensslin2020}, leading to the currently most accurate map of the Galactic Faraday sky in \citet{hutschenreuteretal2022} (Fig.~\ref{f:hutsch22}).

%--------------------------------------
%--------------------------------------

\begin{figure}[ht]
    \centering
    \includegraphics[width=0.85\textwidth]{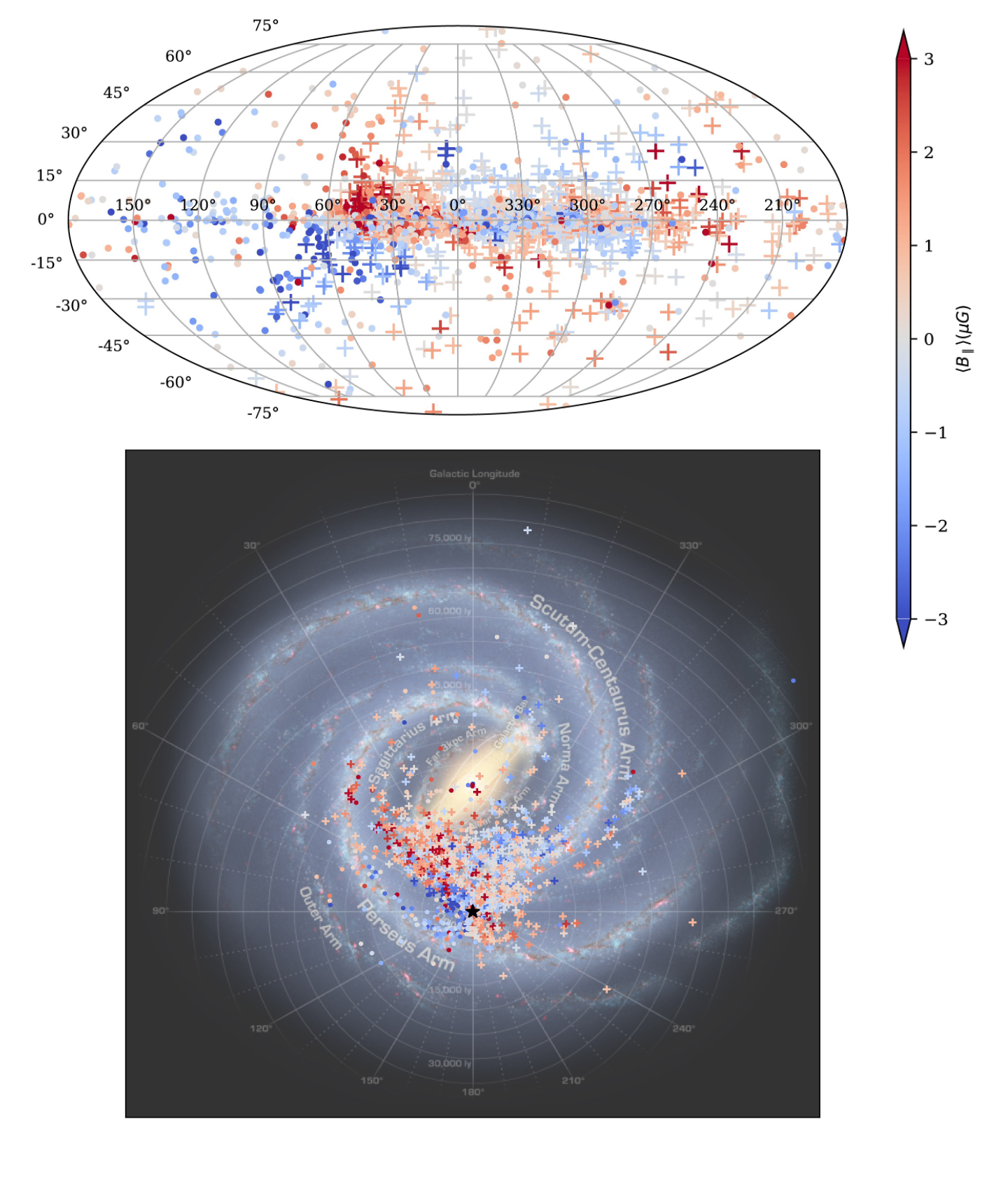}
    \caption{Top: distribution of pulsar RMs on the sky. Color indicates derived magnetic field strength along the line of sight ($B_{\parallel}$), where red (blue) indicates $B_{\parallel}$ towards (away from) the observer, capped at $\pm 3~\mu$G. Dots are pulsars from the {\sc psrcat} catalog, pluses are from \citet{oswaldetal2025}. Bottom: RM distribution of pulsars from the same two sources, only for pulsars with a distance to the plane $|z| < 1$~kpc, overlaid on an artist impression of a bird's-eye view of the Galaxy based on spiral arms, made by Robert Hurt in \citet{churchwelletal2009}. Figures reproduced from \citet{oswaldetal2025} (CC BY 4.0).}
    \label{f:pulsarRM}
\end{figure}

Pulsar RM’s can be used to create a 3D RM grid as long as pulsar distances are known. Known pulsar distances are a great advantage in disentangling magnetic structure along a line of sight, as is the absence of intrinsic or intergalactic RM contributions \citep{noutsosetal2008,hanetal2018,sobeyetal2019,ngetal2020}. Disadvantages of using pulsars is their relatively small number (as compared to extragalactic radio sources), especially outside the Galactic disk, and at times large distance uncertainties. The current largest catalogue of pulsar RMs (as assembled from {\sc psrcat}\footnote{\url{http://www.atnf.csiro.au/research/pulsar/psrcat}} \citet{manchesteretal2005} and \citet{oswaldetal2025}) is shown in Fig.~\ref{f:pulsarRM}. 

From pulsar RMs and DMs one can estimate the line-of-sight averaged strength of the magnetic field component parallel to the line of sight $\langle B_{\parallel}\rangle$ as
\begin{equation}
\langle B_{\parallel} \rangle = 1.232\frac{\mbox{RM}}{\mbox{DM}},
\label{e:dmrm}
\end{equation}
\citep{smith1968,sobeyetal2019}. This approximation holds under the assumption that magnetic field and electron density are uncorrelated, which is reasonable for subsonic and transonic media \citep{setafederrath2021}. However, the average value of the magnetic field strength can be misleading or not very meaningful if there are many reversals in magnetic field direction along the line of sight. For any two pulsars in a similar direction at distances $D_1$ and $D_2$, this quantity can be evaluated along a shorter path length $\mid D_2-D_1\mid$ as \citep[e.g.,][]{hanetal2018}
\[
\langle B_{\parallel} \rangle_{(D_2-D_1)} \approx 1.232\frac{\mbox{RM}_2-\mbox{RM}_1}{\mbox{DM}_2-\mbox{DM}_1}.
\]

%--------------------------------------
%--------------------------------------

An exciting prospect is building an RM grid from Fast Radio Bursts (FRBs) to probe Galactic magnetic fields \citep{pandhietal2022}. FRBs are extremely short (millisecond) bursts of radio emission discovered by \citet{lorimeretal2007}. Although their exact source population is still unknown \citep{zhang2020}, their large DMs indicate that they are extragalactic. Due to their short timescales, both DM and RM can be determined, so that an estimate of their line-of-sight integrated parallel magnetic field strength $\langle B_{\parallel}\rangle$ can be estimated using Eq.~\eqref{e:dmrm}. Observations of FRB RMs and DMs are rapidly growing with the new generation of high-time-resolution radio telescopes \citep{chimefrbcollab2021,hacksteinetal2020}.

%https://ui.adsabs.harvard.edu/abs/2022MNRAS.516.4739P/abstract "A method for reconstructing the Galactic magnetic field using dispersion of fast radio bursts and Faraday rotation of radio galaxies" did simulations of large distribution of FRBs and shows that IFT-based inference can constrain models. Current distribution of FRBs is too small to constrain anything.

%%%%%%%%%%%%%%%%%%%%%%%%%%%%%%%%%%%%%%%%
\subsection{Rotation Measure Synthesis/Faraday Tomography}
\label{s:fartom}

Rotation Measure Synthesis describes the process that generates a Faraday Dispersion Function (FDF), generally called \emph{Faraday spectrum}, from measurements of Faraday-rotated linearly polarized emission at a range of closely spaced frequency channels. The principle of Rotation Measure Synthesis is first explained in \citet{burn1966} and first applied in \citet{bdb2005}. Clear explanations can be found in \citet{heald2009} and \citet{ideguchietal2018}. 

The complex linear polarization as a function of wavelength squared, $P(\lambda^2) = Q(\lambda^2) + iU(\lambda^2)$, can be written as the integral over all components of Faraday depth $\phi$ along a line of sight, i.e.\ a Faraday spectrum $F(\phi)$, as:
\begin{equation}
    P(\lambda^2) = \int^{\infty}_{-\infty} F(\phi) \mbox{e}^{2i\phi\lambda^2} d\phi
\end{equation}
\citep{burn1966}. As the polarized radiation can only be observed over a finite frequency range, the observed polarization $\tilde{P}(\lambda^2)$ can be described as
\begin{equation}
    \tilde{P}(\lambda^2) = W(\lambda^2)\int^{\infty}_{-\infty} F(\phi) \mbox{e}^{2i\phi\lambda^2} d\phi
\label{e:pobs}
\end{equation}
with $W(\lambda^2)$ the window function describing the observed frequency range. Eq.~\eqref{e:pobs} can be Fourier transformed to
\begin{eqnarray}
    \tilde{F}(\phi) &=& F(\phi) * R(\phi) = K \int^{\infty}_{-\infty} \tilde{P}(\lambda^2)\mbox{e}^{-2i\phi\lambda^2}d\lambda^2 \;\;\; \mbox{    where} \label{e:phitilde}\\
    R(\phi) &=& \frac{\int^{\infty}_{-\infty}W(\lambda^2)\mbox{e}^{-2i\phi\lambda^2}d\lambda^2}{\int^{\infty}_{-\infty}W(\lambda^2)d\lambda^2}
\end{eqnarray}
Here, $\tilde{F}(\phi)$ is an approximate reconstruction of the Faraday spectrum $F(\phi)$, considering the limited frequency range, and the \emph{Rotation Measure Spread Function (RMSF)}\footnote{\citet{bdb2005} call this the Rotation Measure Transfer Function (RMTF).} $R(\phi)$ is the normalized Fourier Transform of the window function \citep{bdb2005}. Therefore, observations of polarized radiation at a range of frequencies allow the calculation of a Faraday spectrum, which is convolved with an RMSF that reflects a limited wavelength coverage. The deconvolution of a Faraday spectrum from the RMSF can be attempted using the technique of RM-CLEAN \citep{heald2009,belletal2013}, analogous to the well-known iterative subtraction algorithm CLEAN for radio interferometry \citep{hogbom1974}. 

Faraday Tomography is commonly defined as the application of Rotation Measure Synthesis to frequency cubes, resulting in a Faraday cube consisting of sky maps of polarized intensity at a range of Faraday depths. A Faraday spectrum which consists of a single unresolved peak, corresponding to a Faraday-rotating screen in front of a synchrotron-emission source, is called Faraday thin or Faraday simple. A Faraday thick, or Faraday complex, medium contains Faraday rotation and synchrotron emission interspersed, resulting in a complex Faraday spectrum with multiple and/or broad peaks.

Entirely similar to radio interferometry, the quality of the  reconstruction of the Faraday spectrum, $\tilde{F}(\phi)$, depends on the coverage in $\lambda^2$. As the wavelength coverage is both finite and positive, the Fourier Transform is necessarily based on incomplete information. Therefore, the inversion problem is ill-posed, and additional assumptions have to be made to convert linear polarization into the Faraday spectrum. 

These assumptions could be constraints on the Faraday spectrum (e.g.\ \citet{bdb2005} discuss the situation where the Faraday spectrum is real) or the morphology of the probed magnetic fields (e.g.\ \citet{fricketal2011} assume an even and symmetric magnetic field in face-on disk galaxies). In more complex situations of multiple components and Faraday depth rising non-monotonically with distance, \citet{fricketal2010} introduce Faraday Tomography based on wavelets, which they show can reproduce simulated Faraday depth structures at different physical scales. Furthermore, some artifacts can be avoided by directly applying a 3D Fourier Transform from visibilities to Faraday depth, avoiding the spectro-polarimetric imaging stage in between. \citet{bellensslin2012} pioneered this method, named Faraday Synthesis, and \citet{gustafssonetal2025} show that it, combined with direction-dependent calibration, can increase dynamic range by reducing artifacts.

A much-used assumption to impose upon the unknown Faraday spectrum is the assumption of sparsity, i.e.\ the assumption that the Faraday spectrum consists of only a low number of non-zero components. The signal processing technique using the assumption of sparsity, Sparse sampling or Compressive Sampling (CS) \citep{candesetal2005, donoho2006} seeks a solution which has the smallest number of (thin or thick) Faraday depth components needed to fit the data \citep{lietal2011, andrecutetal2012, akiyamaetal2018, carcamoetal2022}. A computationally relatively inexpensive way to solve the inversion problem is a technique called Constraining and Restoring iterative Algorithm for Faraday Tomography \citep[CRAFT,][]{coorayetal2021}. CRAFT approximates solutions, using iteratively improving assumptions about the extremes of Faraday depth bounds.

If some characteristics of a Faraday spectrum are known or can be estimated, such as the number of components in the spectrum or the wavelength dependent depolarization behavior, one can use parametrized modeling and obtain best-fit parameters for the Faraday components. This fitting, commonly called QU-fitting, was initially performed using a least-squares approach \citep{farnsworthetal2011} or maximum likelihood methods \citep{osullivanetal2012,ideguchietal2014}. Later, these were replaced by more sophisticated fitting methods like  Markov Chain Monte Carlo approaches \citep{ozawaetal2015, kaczmareketal2017, sakemietal2018}, or using Convolutional Neural Networks \citep{brownetal2019}. 

\citet{sunetal2015} make a comparison between various methods to obtain reliable Faraday spectra from Faraday Tomography, viz.\ Compressive Sensing techniques, wavelet transformations and QU-fitting. They found that all methods have difficulty reliably retrieving Faraday-thick structures (see also \citealt{miyashitaetal2019}), but QU-fitting performs better than Fourier-based methods for multiple Faraday-thin components. 

%https://arxiv.org/pdf/2102.10903.pdf - quick FT complexity classification by Alger, Livingston, NMc-G

%%%%%%%%%%%%%%%%%%%%%%%%%%%%%%%%%%%%%%%%
\subsection{Gradient techniques}
\label{s:gradients}

%Schmaltz, Yu, Lazarian (2024) on SF region L1688: in inner Galaxy, don't realize distance ambiguity and use HI for velocity slices and rotation curve --> wrong. Other problems with this paper are not realizing that pol angle 180 is equal to 0, and possibly using rotation curve to make slices in CO (or possibly CO single slice). 
%Hu & Lazarian (2023): outer galaxy, velocity misinterpreted. 
%Hu \& Lazarian (2023 July) -- trouble with cloud distances, method OK but cloud distances are not. So not applicable to outer Galaxy.

Magneto-hydrodynamic (MHD) turbulence is anisotropic, with eddies elongated along the local magnetic field orientation \citep{goldreichsridhar1995}. Therefore, one would expect turbulent fluctuations to be aligned with the local magnetic field orientations. In addition, magnetic flux freezing causes gas motions along field lines, resulting in gaseous filamentary structures aligned with the local magnetic field. In principle, any observable that traces these anisotropic fluctuations can be used to probe the local magnetic field orientations, without the need for accurate spectro-polarimetric data. Fluctuations in velocity, velocity centroids, synchrotron intensity, synchrotron polarization, dust emission or dust polarization have been used to analyze magnetic fields, of which an overview follows:
\begin{itemize}
%item  \citep{kowaletal2007}: statistical properties of density / spectra differ for strong and weak fields. Not really
\item \textbf{Radio Polarization Gradients} \citep{gaensleretal2011} are spatial gradients of polarization pseudo-vectors. Guided by turbulence properties of these gradients as determined from MHD simulations \citep{burkhartetal2012}, \citet{gaensleretal2011} show that the properties of the distributions of these gradients can indicate whether the magnetized turbulence is subsonic, transonic or supersonic.
\item \textbf{Gradients in dust continuum emission} correlate with orientations of dust polarization in molecular cloud cores. Using the MHD force equation, \citet{kochetal2012} present a method to calculate magnetic field strengths in these cores from the angle difference between emission gradients and polarization orientation, if the local gravitational forces and pressure gradients can be estimated. Even though \citet{kochetal2012} applies the method to a collapsing cloud core, they state that the method is applicable more broadly, as long as local gravitational forces and pressure gradients can be determined.
\item \textbf{Histograms of Relative Orientations} (HRO, \cite{soleretal2013}) compare plane-of-sky projected magnetic field orientations to gradient directions in density structure in molecular clouds. \citet{soleretal2013} show in simulations of self-gravitating molecular clouds that the orientations of magnetic field and density gradients change from aligned to perpendicular depending on density and magnetization of the clouds. \citet{huetal2019} modified the HRO algorithm to use velocity instead of density structures.
\item \textbf{Velocity Gradient Technique} (VGT): gradients in the velocity field as a tracer of magnetic fields were introduced by \citet{gonzalez-casanovalazarian2017}, who showed that in numerical simulations, for sub-Alfv\'enic turbulence, velocity gradients indeed correlate with local magnetic field orientations. This method makes it possible to trace the plane-of-sky magnetic field orientations from H~{\sc i} spectra in absence of polarimetric data, and avoids some of the complexities present in density fluctuations analysis. Velocity gradients have the potential to map out the plane-of-sky magnetic field orientation in three spatial dimensions if velocity information can be reliably mapped into distance information, which is possible for part of the Galactic plane. In addition, \citet{lazarianetal2018_vgt} showed that the velocity gradient distribution is related to the Alv\'enic Mach number in MHD simulations, and therefore allows estimation of the magnetic field strength.
%applied to H I data (Lazarian & Yuen 2018b) and molecular clouds (Hu et al. 2019a).
\item \textbf{Synchrotron Intensity Gradients} (SIGs), introduced in \citet{lazarianetal2017}, use gradients in synchrotron intensity as proxies for line-of-sight integrated magnetic field orientation. Advantages of this method are that synchrotron intensity is strong at lower wavelengths, and that it does not contain the complexity of Faraday rotation. The method is not applicable in the Galactic plane, as synchrotron structures there tend to not be associated with interstellar turbulence (only). Comparison of Planck synchrotron intensity gradients with Planck synchrotron polarization angles shows that at high latitudes, gradients agree well with polarization angles in about half of the sky (Fig.~10 in \citet{lazarianetal2017}). Simulations show better alignment in subsonic turbulence than in supersonic turbulence \citep{holazarian2023}. \citet{kochetal2012} conjecture that their method (presented above) is also applicable to gradients in synchrotron intensity under certain conditions. 
% Lazarian & Pogosyan (2012): theory of relations between synchrotron fluctuations and underlying turbulence. INCLUDE OR NOT?
\item \textbf{Synchrotron Polarization Gradients} (SPGs, \citet{lazarianyuen2018_sgp}), defined as gradients in polarized intensity $P$, align with line-of-sight averaged magnetic field significantly when averaged over large scales, but less so on smaller scales. In the limit of low Faraday rotation, its results are similar to SIGs, as expected. %The same paper also investigates gradients in the $\lambda^2$-derivative of complex polarization in the limit of low Faraday rotation, which gives similar results on tracing of magnetic field. 
% Ho, Yuen, Leung, Lazarian (2019) compare SPGs with Faraday tomography and conclude that SPGs still work if FT has eg too small frequency coverage. --> cited later.
SPGs retrieve magnetic field parameters consistent with power spectra of polarized intensity in spectro-polarimetric radio data \citep{zhangliu2025}, and to Faraday Tomography based methods, tracing magnetic fields also in regimes where Faraday Tomography is not reliable \citep{hoetal2019}.
\item \textbf{Intensity Gradients} \citep[IGs,][]{huetal2019} are defined as gradients in intensity in velocity-channel-integrated maps of neutral hydrogen (or other species), which trace the gas density fluctuations and not velocity fluctuations. Numerical simulations show that velocity gradients are more accurate and reliable predictors of magnetic field than intensity gradients, although intensity gradients trace high density contrasts around shocks and can therefore be used to identify shocks.
% OTHER GRADIENTS:
% Dust Polarization Gradients (DPGs) were also discussed in Lazarian & Yuen (2018a). 
% gradient of gradient amplitude (GGA), Ho & Lazarian (2021) "In Ho & Lazarian (2021) we proposed (...) GGA, which improves the magnetic field tracing by gradients. However, an in-depth study is required to analyse the applicability of GGA in subsonic regimes versus the change of Alfv ´en Mach number"
\end{itemize}

\section{The coherent Galactic magnetic field component}
\label{s:coherent}

The galaxy-scale, coherent component of the Milky Way's magnetic field is determined through modeling constrained by measurements of mostly synchrotron emission, Faraday rotation, polarized emission from dust and/or interstellar polarization of starlight. Generally, two kinds of models are being used: parametric heuristic models and parametric dynamo-based models. Non-parametric models of the Galactic magnetic field include magnetic fields on all scales; these show great promise but are for the moment still confined to small parts of the Galaxy, therefore mostly probing magnetic fields on relatively small scales. 

Parametric heuristic models generally piece together a magnetic field structure from log-normal spiral arms and other magnetic field components. Apart from commonly being divergence-free, there is little or no physically rigid basis to the magnetic field structures in these models. Advanced parametric heuristic models from the 2010s and earlier are based on an axisymmetric spiral structure in the Galactic disk, often with exponential decay away from the disk. Various models include vertical magnetic field structure in the halo; some include the turbulent component of the magnetic field addressed in various ways, see the review by \citet{jaffe2019}. Since that review, some advanced heuristic models have been published: \citet{ungerfarrar2023} consists of a suite of eight different models, and \citet{korochkinetal2025} includes modeling of the Local Bubble for the first time (see Sect.~\ref{s:lb}). Both models are fit to Galactic RMs from extragalactic sources, and polarized synchrotron emission from the WMAP and Planck satellites. \citet{beckerkachelriess2025} provide a toy model to explain both synchrotron polarization and cosmic ray diffusion data, including a disk, halo, and extended halo component.
Even though these models have been used extensively and successfully for Galactic and extragalactic research, they have limitations. Apart from using idealized spiral structure (see Sect.~\ref{s:coherent-disk}) and simplified treatment of small-scale fields (see Sect.~\ref{s:smallscale}), they rely on necessarily incomplete and uncertain models for cosmic ray electrons, thermal electrons, and (for some of these) dust distributions; also, there is little connection with the underlying physical processes.

% Korochkin: spiral arms, the toroidal halo, the X-shaped field, and the field of the Local Bubble, using Faraday rotation measures (RMs) of extragalactic sources, together with the synchrotron polarisation data from WMAP and Planck
% Henriksen (2022) shows that X-fields can be created by a disk field, winds and rotation

An alternative approach to these heuristic models is to include dynamo physics in Galactic magnetic field models. Analytic models of such dynamo-based galactic magnetic field configurations are described e.g.\ by \citet{shukurovetal2019} who derive magnetic field models based on eigenfunctions of the mean-field dynamo equations or of the magnetic field diffusion equation, or \citet{ferriereterral2014} who present a number of X-shaped halo magnetic field models. The latter models are used to show that the Galactic magnetic field is possibly bisymmetric and X-shaped \citep{terralferriere2017}, and to model halo fields in otherwise heuristic models \citep{ungerfarrar2023}. 
An ambitious project with the aim of reconstructing the Galactic magnetic field through both parametric modeling and non-parametric inversion methods is the Interstellar MAGnetic field Inference Engine (IMAGINE, \citet{boulangeretal2018,steiningeretal2018}) project.
%Zhang, Green et al 2025 do inversion method on extinction curve to determine 3D RV map.

%\citet{terralferriere2017} compared RMs of extragalactic sources with analytical models of X-shaped magnetic fields in the galactic halo from \citet{terralferriere201} and concluded a slightly higher probability for a bisymmetric than an axisymmetric field. 
%Henriksen dynamo models mostly for external spirals/CHANG-ES. Applied to MW in West helicity paper.

% TOO MUCH DETAIL ON UNGERFARRAR
%Parametric heuristic models steadily increase in number and complexity of components, enabled by inclusion of more and more diverse observational data. The latest suite of models by \citet{ungerfarrar2024} contains a (spiral) disk component, a toroidal (azimuthal) field component and a poloidal field component, all modeled in various ways. They fit their models to almost 45 000 extragalactic point source RMs averaged over $\sim 3.66^{\circ}$ sky pixels, and an all-sky Stokes $Q$ and $U$ map at 30~GHz derived from WMAP 22.5~GHz and Planck 30~GHz measurements. %containing a poloidal field component created by the Milky Way's differential rotation. 

Evidence accumulates that the large-scale magnetic field in the Galactic disk behaves differently from that in the Galactic halo. Therefore, we will discuss these two components separately in the two following subsections.

%Huge body of dynamo models as summarized by Brandenburg (2015)
%Also more MHd/dynamo models
%- Reissl
%- Gressel et al 2013 undulating field reversals in the disk. 
%- Dobbs et al 2016 also describe large-scale spirals, cauased by high velocity jumps %across the spiral shock and/or at co-rotation.  (SPMHD simulations)
%- pakmor et al (2018) find RMs crudely consistent with observed RMs in the Galactic disk, but warn that additional structure depends strongly on the position of observer and local environment.

%Wu, Kim, Ryu (2015) show significant correlations between ne and B. Unger \& Farrar allow significant correlation (not anti-correlation) but can model with correlation 0 as well as long as you introduce striated field. 

%Han et al (2018) uses pulsar RMs and distances to infer a magnetic field model in the disk with independent magnetic field values in arms and interarms. He finds reversals in every arm, but does not quote errors and does not give any details about how the fitting is done. Are magnetic field values fitted freely, or are directions imposed? CHECK WITH HAN.

%%%%%%%%%%%%%%%%%%%%%%%%%%%%%%%%%%%%%%%%
\subsection{Coherent magnetic fields in the Galactic disk}
\label{s:coherent-disk}
The coherent disk magnetic field follows the spiral arms to first order. This is  observed in any nearby, (partially) face-on external spiral galaxy, although  deviations from a regular spiral pattern exist \citep{beckwielebinski2013}. Milky Way data also strongly suggest magnetic field orientations following material spiral arms, at least approximately.
%jaffe2019 review radiopol; heiles1996 optical pol; other refs?
Although an even-parity (log-normal) spiral structure is the best starting point for parametric modeling of large-scale magnetic fields in or close to the Galactic plane, there is no denying that this description is wildly oversimplified.
It is still unclear how well spiral arms are followed, and how the field strength varies in spiral arm and interarm regions. There is strong evidence for (at least) one reversal in the magnetic field direction on scales of kiloparsecs or more, but no agreement on the exact location, number, and extent of magnetic field reversals.

%%%%%%%%%%%%%%%%%%%%%%%%%%%%%%%%%%%%%%%%
\subsubsection{Field strength}
Synchrotron intensity measurements indicate a total Galactic magnetic field strength in the Solar neighborhood of $6\pm2~\mu$G, increasing towards the Galactic Center to about $10\pm3~\mu$G at 3~kpc distance \citep{beck2001}. These calculations assume equipartition, which is likely valid at large scales \citep{setafederrath2021}. The coherent part of this field is estimated from pulsar data through Eq.~\eqref{e:dmrm} to be $1.4\pm0.2~\mu$G in the Solar neighborhood, increasing towards the Galactic Center to $4.4\pm0.9~\mu$G in the Norma arm, assuming uncorrelated magnetic field and thermal electron density \citep{beckwielebinski2013, hanetal2006}. These values are qualitatively supported by global parametric models. Those models that include random components tend to show turbulent fields dominating over coherent fields, with comparable isotropic and anisotropic components \citep{jaffeetal2010,janssonfarrar2012b,orlandostrong2013}. \citet{beckerkachelriess2025} argue that random fields dominate over regular fields only in the Galactic disk but not in the halo. This is also qualitatively consistent with face-on external spirals, where ratios of isotropic turbulent to ordered (coherent and anisotropic random) fields are observed to be more than 5 in spiral arms, and 0.5--2 in interarm regions  \citep{beckwielebinski2013}. Fluctuation dynamos would produce ratios of $B_\mathrm{coh}/B_\mathrm{ran} \sim 0.03 - 0.1$, and tangling of coherent magnetic fields by turbulent gas or interstellar shocks is expected to result in comparable coherent and random field strengths (SS21). 

%jaffe+10: "ratios of the field components’ energy densities that we measure as roughly 1:5:4 (coherent:random:ordered)." So field strengths 1:sqrt(5):2
%ungerfarrar24: only coherent no turbulent component
%janssonfarrar2012b: fit for a striation factor (B_stri/B_reg)^2 = 2.9, but degenerate with n_cre

%%%%%%B-n_e relation
The magnetic field strength $B$ depends on the interstellar density $n$ for both dense (molecular and atomic) and diffuse (atomic and ionized) gas, often modeled as a power law $B \propto n^\alpha$ with different spectral indices $\alpha$ for dense and diffuse gas. An early, comprehensive study of Zeeman splitting in molecular clouds and diffuse HI clouds found that $\alpha \approx 2/3$ for $n \gtrsim 200$~cm$^{-3}$, while $B$ was consistent with independent of $n$ at lower densities \citep{crutcheretal2010}. Subsequent studies, including ionized gas by way of pulsar measurements and/or DCF measurements in neutral clouds, tend to find similar broken power laws, mostly with also non-zero (slightly positive) dependencies of $B$ on $n$ for the diffuse gas (\citealt{harveysmithetal2011,tritsisetal2015,kalberlahaud2023}, however, see \citealt{jiangetal2020}).
Recasting the $B-n$ relation into the dependence of magnetic energy on turbulent kinetic energy, \citet{setamccluregriffiths2025} showed that this satisfies a single relation $E_\mathrm{mag} \propto E_\mathrm{kin}^\beta$, with $\beta \approx 0.64 - 0.72$ (depending on the data from which turbulent velocities are derived). This relation can be physically understood as a fraction of the kinetic energy converted to magnetic energy. They also conclude that magnetic field fluctuations are caused by both density and velocity fluctuations, and that magnetic and thermal pressures are comparable in all ISM phases.

Many spiral galaxies show stronger ordered magnetic fields in the material interarm regions than in the spiral arms, sometimes ordered into 'magnetic arms'. This may be accompanied by enhanced magnetic fields at the inner edges of the material arms, or magnetic fields crossing over spiral arms  \citep{patrikeevetal2006,beck2015}. Determining the relative field strengths in spiral arms vs interarm regions in the Milky Way is more difficult due to our inside vantage point, as evidenced by varying results from GMF models. 
\citet{jaffeetal2013} find large coherent and anisotropic random magnetic field strengths co-located with dust arms, but an enhanced isotropic random field strength in the interarm regions. This can be caused by a large-scale shockwave associated with the arms that compresses isotropic random fields, also consistent with varying RM structure functions in Galactic arms and interarm regions \citep{haverkornetal2006,haverkornetal2008}. This is confirmed by a pulsar study in the first and second Galactic quadrant, which finds stronger large-scale magnetic fields in the arms than in the interarms \citep{curtinetal2024}. By contrast, \citet{ungerfarrar2023} find a minimum coherent field strength at the material arm locations, and a maximum coherent field strength in the interarm regions, indicating 'magnetic arms' as in some external spirals. 

% Poldermanetal
Low-frequency absorption of HII regions indicates that surplus synchrotron emissivity exists in the far Galactic plane that is not accounted for in the GMF models; either caused by enhanced magnetic field strengths, meso-scale magnetic field orientation changes, or an overdensity in Galactic cosmic rays \citep{polderman2019,polderman2020}.

\subsubsection{Orientation of the disk magnetic field}
\label{ss:orientationdisk}

Although the Galactic magnetic field roughly follows the spiral arm directions in the Galactic disk, detailed studies reveal meso-scale structures in the orientation of the magnetic field in various ways.

 %Xu \& Han (2019) discuss an anomalous region of reversed magnetic field (also in Han et al 1999, 2006, the Carina anomaly). The region has many HII regions and enhanced DM and RM. Change in magnetic field direction due to "a component of the random magnetic field or a systematic distortion of the uniform field"
%asymmetry in north and south, Curtin, ordog 2017, M et al 2020. Curtin: possibly due to combination of didks and halo field (even/odd)

\paragraph{Pitch angle} 
The Milky Way's material arms have a pitch angle\footnote{The pitch angle $p$ of a spiral magnetic field is defined as $\tan p = \overline{B_r}/\overline{B_\phi}$, where $B_r$ is the radial component of the magnetic field and $B_\phi$ its azimuthal component.} $p \sim11^{\circ}$ \citep{reidetal2019, houhan2014}. Many parametric GMF models either use this pitch angle value as an input parameter, or find similar pitch angles when fitting data. However, pitch angles $2-3$ times as high or as low have been reported, see Table~1 and its discussion in \citet{haverkorn2015} for more background.

Standard non-linear mean-field dynamo models in a thin disk predict pitch angles $p \lesssim -20\dg$, although \citet{chamandytaylor2015} show that variations in disk scale height or turbulence correlation length can result in widely varying pitch angle values. The prediction that the pitch angle decreases with distance to the Galactic centre is tentatively observed in some nearby spiral galaxies (SS21, Fig 13.1), but this possibility is not yet included in any of the heuristic magnetic field models in the Milky Way.

Local derivations of the pitch angle have been observed e.g.\ towards the Galactic anti-center, where the local magnetic field direction in the Perseus arm is most consistent with a pitch angle of $\sim0\dg$ \citep{raebrown2010,vanecketal2011}.

\paragraph{Large-scale reversals in magnetic field direction}
At the Sun's location, the coherent magnetic field is oriented along the spiral arms, and directed clockwise when seen from the North Galactic pole. There has been strong evidence for many decades for a reversal in this field direction towards the inner Galaxy, directed along a spiral arm \citep{thomsonnelson1980,simardnormandinkronberg1980}. %,hanqiao1994,randlyne1994}. Cite only base article
This observation started the ongoing discussion on large-scale Galactic reversals in magnetic field direction along spiral arms. 
Pulsar RMs are in many ways ideal probes for spiral arm reversals, as they are mostly located close to the Galactic plane and many pulsars have known distances. However, interpretation of the results requires care as these studies rely on thermal electron density models, generally assume a certain (or no) correlation between electron density and magnetic field, and use at times very uncertain distance estimates. Pulsar studies relying on differential RM and DM measurements along lines of sight (often combined with extragalactic source RMs) tend to find many reversals in the coherent field direction, from arm to arm or even between every arm and interarm, both in the inner and the outer Galaxy \citep{%hanetal1997,hanetal2006,
hanetal2018,xuetal2022}.
% only cite two most recent
% ook han et al 2002: detection of reversal in Norma arm
\citet{curtinetal2024} introduce a new method to determine spiral arm magnetic fields from pulsar data in the northern sky.
Using Eq.~\eqref{e:dmrm}, they calculate a magnetic field which is assumed to be directed along a spiral arm and constant in strength, per spiral arm or interarm region. Then they 'geometrically correct' the RM by subtracting the resulting longitude-dependent RM in any foreground arms and interarms from further pulsar RMs. This results in magnetic field strengths and directions in spiral arms and interarm regions, at both sides of the Galactic midplane. They confirm the well-documented reversal between the Sagittarius and Perseus arms (agreeing with \citet{hanetal2018}), but do not find a reversal in the Perseus arm (disagreeing with \citet{hanetal2018}).

%https://arxiv.org/pdf/2101.05384 Seta and Federrath: Bpar = 1.232RM/DM is OK on large scales, but breaks down on small scales

Parametric Galactic magnetic field models of log-normal spiral arms fitted to extragalactic source RMs (often combined with pulsar RM measurements) allow global modeling out to larger distances, but these RMs are integrated throughout the entire Milky Way as opposed to pulsar RMs. These models find typically less (but not zero) reversals along spiral arms, but do not give a consistent number of reversals \citep{jaffe2019}. 

No global reversals have been observed in external spirals \citep{beckwielebinski2013}, which leads to alternative explanations for the Milky Way data. The field reversal may be only a few kpc in size \citep{ungerfarrar2023}, similar to a local reversal observed in M51 \citep{berkhuijsenetal1997}. Alternatively, the apparent reversals may in fact be caused by local structure such as H~{\sc II} regions \citep{mitraetal2003}. 

The mean-field dynamo mechanism can create global field reversals in galactic disks \citep{ruzmaikinetal1985}, but can also result in local reversals at approximately the Galactocentric radius where the nearest observed magnetic field direction reversal is found \citep{bykovetal1997}. MHD simulations of spiral galaxies show magnetic field reversals for some spiral potentials, in locations of large velocity changes across spiral shocks \citep{dobbsetal2016}.
\paragraph{The region around the Sagittarius arm tangent}
\label{s:sgrarm}

\begin{figure}[ht]
    \centering
    \includegraphics[width=\linewidth]{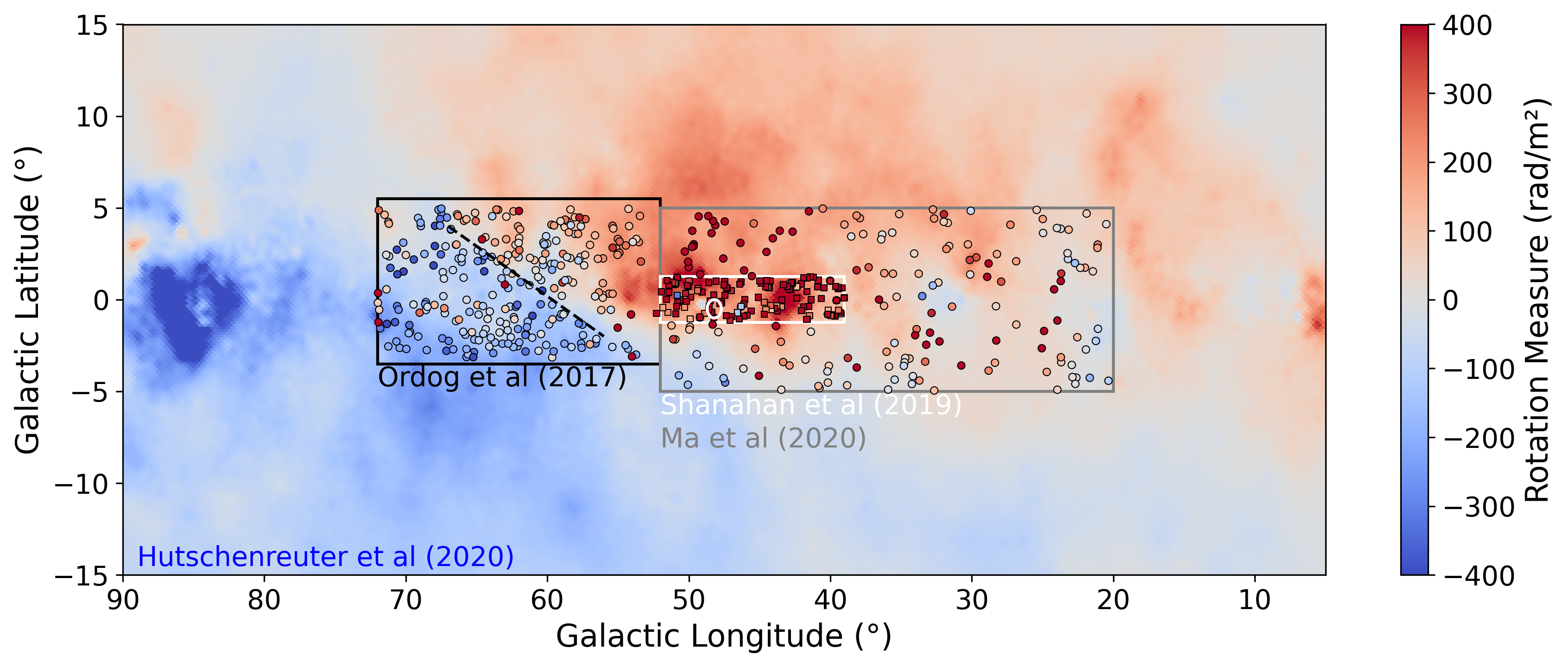}
    \caption{Combination of studies in the Sagittarius arm tangent region. The colored background denotes Galactic RMs from \citet{hutschenreuteretal2022} (corrected for free-free emission), overlaid with extragalactic source RMs from \citet{ordogetal2017} (circles, black box), \citet{maetal2020} (circles, grey box) and \citet{shanahanetal2019} (squares, white box). The black dashed line is a sign reversal discussed in \citet{ordogetal2017}, and the yellow band indicates the approximate region of RM reversal according to the model by \citet{boothetal2026}. The small white ellipse shows the  area of anomalously large RMs ($\gtrsim 1000$~rad~m$^{-2}$) discussed in \citet{shanahanetal2019}.}
    \label{f:sagregion}
\end{figure}

One particularly well-studied area concerning magnetic field reversals is 
the approximate direction of the Sagittarius arm tangent, from $\ell \sim 40\dg$ to $\ell \sim 70\dg$. This region contains a large-scale reversal in sign of extragalactic RMs, which is significantly tilted with respect to the Galactic plane. 
Figure~\ref{f:sagregion} relates the various RM measurements in this region of sky. \citet{ordogetal2017} (black box in Figure~\ref{f:sagregion}) noted that this reversal was present both in extragalactic source RMs and in Faraday depth measurements from diffuse extended synchrotron emission (with an estimated polarization horizon of $\sim2$~kpc), with the extragalactic source RMs being $\sim 150$~rad~m$^{-2}$ larger in value. \citet{ordogetal2017} offer various explanations for this consistency in sign but difference in value of the RMs: a larger polarization horizon than assumed, a quick decay in magnetic field and/or electron density with galactic radius, or RM sign reversals at large distances.

\citet{boothetal2026} proposed that this RM structure is due to a local ($\lesssim 550$~pc) magnetic field reversal which is slanted with respect to the Galactic plane and which passes the plane between the Local and Sagittarius Arms. Their model is local and therefore considers a simple planar reversal in a homogeneous medium, but recovers the large-scale RM structures remarkably well. This model implies that the local medium dominates the Faraday sky except in a narrow Galactic disk region, as also indicated in MHD simulations of Milky Way-like galaxies \citep{pakmoretal2018}.

%ordog
%\citet{ordogetal2017} focused on a sign reversal in RM towards the Sagittarius-Carina arm, along a \emph{diagonal} in the plane of the sky, between Galactic coordinates $55^{\circ} \lesssim \ell \lesssim 67^{\circ}$ and $-2^{\circ} \lesssim b \lesssim 4^{\circ}$. Figure~\ref{f:sagregion} shows the extragalactic RMs in their field, with a diagonal dashed line at the location of the sign reversal. They interpret this as a slanted reversal in the large-scale magnetic field, inclined in the plane of the sky, and possibly along the line of sight as well.  This reversal is visible in both Faraday depth from diffuse extended synchrotron emission (with an estimated polarization horizon of $\sim2$~kpc) and from extragalactic point sources, with the extragalactic source RMs being $\sim 150$~rad~m$^{-2}$ larger in value - a small difference considering the ratio of the polarization horizon to the path length through the Galaxy. \citet{ordogetal2017} offer various explanations for this consistency in sign but difference in value of the RMs: a larger polarization horizon than assumed, a quick decay in magnetic field and/or electron density with galactic radius, or RM sign reversals at large distances. 

This conclusion is supported by various other observational studies. 
From analysis of new low-latitude pulsar RMs, \citet{oswaldetal2025} conclude that the RM is mostly antisymmetric across the plane, but symmetric in the direction of the Scutum–Centaurus and Sagittarius arms. They suggest that the plane of antisymmetry in magnetic field may be tilted or curved.  
This symmetry across the plane in the Sagittarius arm was also noted by \citet{maetal2020} in extragalactic source RMs. They use models of local odd-parity magnetic field reversals in various spiral arms to explain the RM sign change across the Galactic plane, but their observations seem also consistent with the \citet{boothetal2026} slanted reversal model.
\citet{curtinetal2024} notice asymmetries across the midplane in magnetic field strength, which they also interpret as an inclined magnetic field reversal with respect to the line of sight. \citet{xuhan2019} discuss a magnetic field reversal in this direction at a distance of $\sim 1$~kpc. They report both positive and negative pulsar RMs in this direction at distances of $\gtrsim 1$~kpc, which may well be consistent with a tilted reversal.

\citet{shanahanetal2019} discovered unexpectedly high extragalactic source RM values in the high-RM Galactic plane region as indicated in Fig.~\ref{f:sagregion}.
In particular, RMs over 1000~rad~m$^{-2}$ were found in a few degrees around longitude $48^{\circ}$, interpreted as being due to enhanced electron density at the inside of the Sagittarius spiral arm, possibly coupled to compressed magnetic fields. These anomalously high RMs were not detected in pulsar RMs, leading \citet{curtinetal2024} to conclude that the anomalous RMs were created at distances beyond the Sagittarius arm. However, none of these pulsars are actually located in the small region of anomalous RMs (the elliptical region in Figure~\ref{f:sagregion}), so the high RMs in extragalactic source RMs and their absence in pulsar RMs may still be consistent if the high RMs are concentrated in a very small region (about 100~pc at the distance of the Sagittarius arm). The enhanced electron density can be due to a shock wave located along the edge of the spiral arm, as also detected in external galaxies \citep{shanahanetal2019}.

%Ma et al 2020: small field of EGSs in first quadrant b<5; find degree-scale deviations; reversal along plane. None of GMF models (Sun, VanEck, JF12) fit well, but VanEck fits the best, as makes sense since that is a disk-only model. Explain deviations by: local disk-odd field model; contribution from odd halo field [not likely as only at certain longitudes]; contamination by local ionized structures [not likely as no Halpha]. Adopted VanEck to include odd-parity disk field

%%%%%%%%%%%%%%%%%%%%%%%%%%%%%%%%%%%%%%%%
\subsection{Coherent magnetic fields in the Galactic halo}
\label{s:coherent-halo}

%xshape
Polarized synchrotron emission from nearby edge-on spirals generally indicates a vertical (i.e.\ perpendicular to the galactic midplane) component of magnetic field, extending a few to many kpc away from the galactic disk. Although less straightforward to characterize in the Milky Way, there is ample evidence for a vertical component in the large-scale magnetic field in the Milky Way halo as well. Dynamo models predict various large-scale configurations of the halo magnetic field, which are being constrained with an increasing amount of observations.

\begin{figure}[ht]
    \centering
     \includegraphics[width=\linewidth]{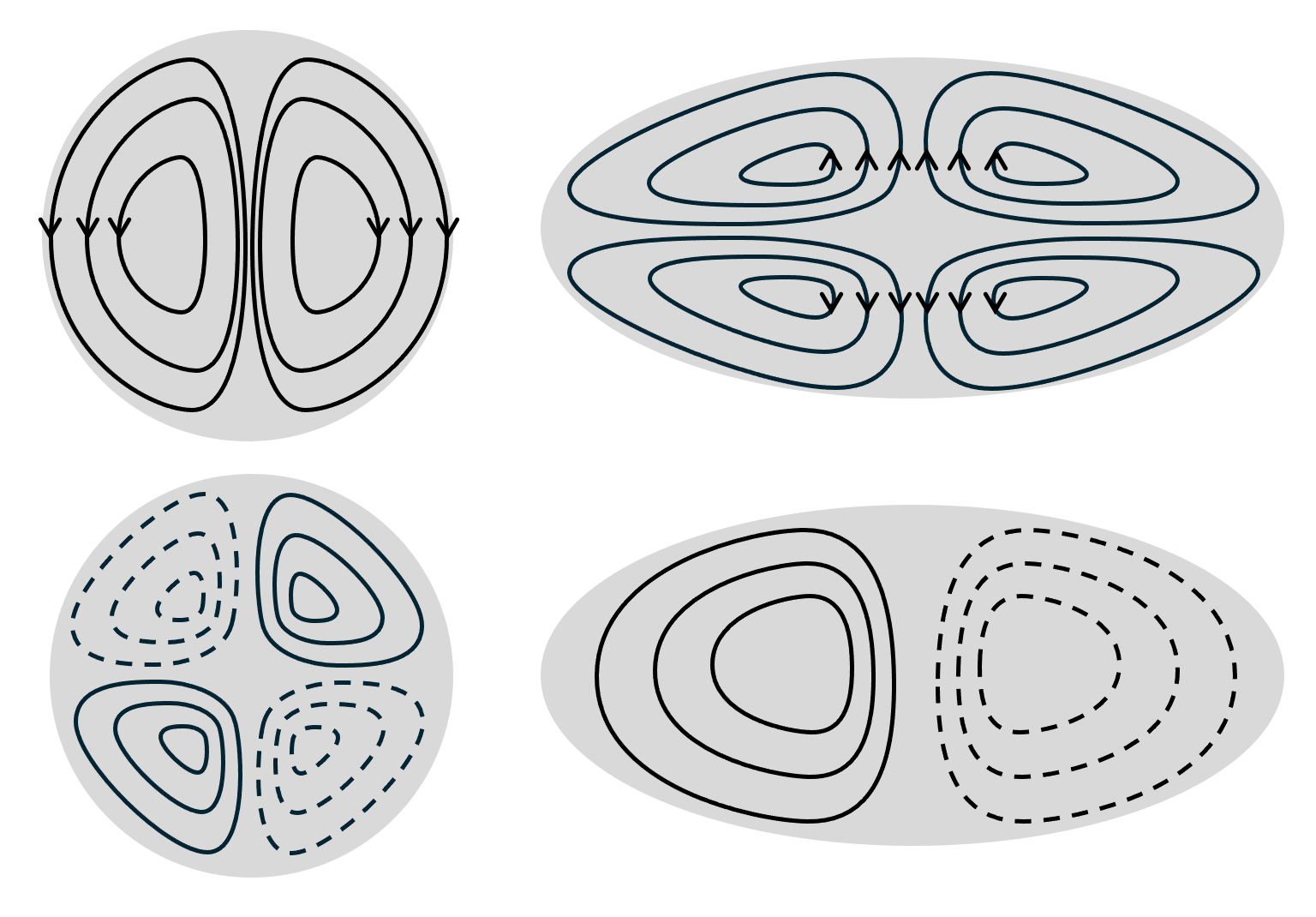}
    \caption{Sketch of field lines in a spherical object where dipolar modes dominate (left), and in a disk-like object where quadrupolar modes are more easily excited (right). Top panels: magnetic field lines of the poloidal field component. Bottom panels: contours of constant field strength of the toroidal field component, where solid and dashed lines denote opposite magnetic field directions into and out of the plane.}
    \label{f:fieldstructures}
\end{figure}

%scale height
The scale height of the magnetic field can be estimated from measurements of rotation measures and dispersion measures of high-latitude pulsars. Fitting exponential scale heights to these data give widely varying results depending on Galactic quadrant and hemisphere, % (from 0.8~kpc to 3.7~kpc), 
with an average scale height over quadrants I and II and northern and southern hemispheres estimated as $2.7\pm0.3$~kpc \citep{xuetal2022} or $2.0\pm0.3$~kpc \citep{sobeyetal2019}. Note that this is effectively a lower limit for the magnetic field scale height due to the measurements' dependence on the electron density scale height.

%butterfly pattern
A conspicuous clue on large-(angular-)scale magnetic structures in the Galactic halo are large anti-symmetries in rotation measure at mid and high Galactic latitudes, known as the 'butterfly pattern', which have been known for many years \citep{simardnormandinkronberg1980}. 
The large-scale patterns on both sides of the Galactic plane are described by 
\citet{dickeyetal2022} as a $\sin(\ell)$ pattern in the Southern sky and a $\sin(2\ell)$ pattern in the Northern sky, with $\ell$ Galactic longitude. They show that these patterns at mid-latitudes are consistent with some variants of dipolar and quadrupolar dynamo models. These structures can similarly be interpreted by invoking a toroidal magnetic field in the Galactic halo, with reversing rotation measure sign above and below the Galactic plane \citep{hanetal1997,prouzasmida2003,xuhan2024}. Alternatively, they can be attributed to magnetized foreground structures such as the Galactic loops \citep{wollebenetal2010, dickeyetal2022}, a magnetic field component in, and along, the Local Arm \citep{ungerfarrar2023}, or a diagonal local magnetic field reversal as discussed in Sect.~\ref{ss:orientationdisk}.

% vertical component
In addition to a toroidal component, there is evidence for a poloidal (often also indicated as vertical or X-shaped) component of the Galactic halo magnetic field. Parametric Galactic magnetic field models including a poloidal component tend to better capture high-latitude structure than models with magnetic field parallel to the disk only \citep{janssonfarrar2012, ungerfarrar2023}. Faraday moment analysis shows that Faraday depths become increasingly positive as a function of $\mathrm{cosec}(b)$ in the northern hemisphere and increasingly negative in the southern hemisphere. This strongly suggests a large-scale vertical magnetic field component pointing from the northern to the southern hemisphere at the Solar radius \citep{dickeyetal2019}, as does analysis of high-latitude extragalactic source RMs in the Solar neighborhood \citep{hanqiao1994, tayloretal2009}. In addition, only Galactic magnetic field models including a vertical component can explain the shape of barrel-like supernova remnants expanding into the local magnetic field \citep{westetal2016}.

%Galactic wind
There is some evidence for a magnetized wind or outflow from the (vicinity of the) Galactic Center, from a large RM gradient within globular Cluster 47Tuc in the direction perpendicular to the Galactic plane, interpreted as arising from interaction with a Galactic wind \citep{abbateetal2020}, and from radio-polarimetric measurements of large radio lobes extending above and below the Galactic Center, similar to the Fermi bubbles \citep{carrettietal2013}.

In the standard paradigm, a dynamo mechanism is invoked to explain the large-scale coherent magnetic field in the Milky Way halo --- although alternative explanations of a vertical component to the halo magnetic field such as a battery mechanism \citep{myserliscontopoulos2021} or magnetohydrodynamic flows due to galactic wind and rotation \citep{henriksen2022} have also been proposed. In a spherical object like a star, dipolar magnetic fields are most easily excited \citep{parker1971}, while a quadrupolar field is more common in a disk-like object \citep{stix1975}, see Fig.~\ref{f:fieldstructures}. This gives credence to the idea that in the Milky Way, combining a flat disk with a more spherical-like halo, a combination of quadrupolar, odd-parity fields in the halo and dipolar, even-parity fields in the disk may be a zeroth-order approximation for the field structure in the Galactic halo \citep{sokoloffshukurov1990}.
However, theoretical conditions for these modes to coincide are specific and time dependent \citep{mosssokoloff2008,ferriereschmitt2000}. Some observational \citep{sunetal2008} and theoretical \citep{shukurovetal2019,ntormousietal2020} evidence exists that such a disk-even-halo-odd configuration may be possible in the Milky Way. However, more evidence points towards a  more complicated halo magnetic field structure that is not consistent with a dipolar or quadrupolar field nor a combination of these \citep{maoetal2010,paveletal2012}.

%Strength of halo field
%"Beck \& Wielebinski: "The field geometry around the Smith High Velocity Cloud, located about 3 kpc below the Galactic plane, requires a surprisingly high strength of the regular halo field of +5 μG (Betti et al. 2019)"

%HVCs:
% Forster, Foster et al (2025): HVCs in HI have some correspondence in polarization. calculate Bpar = 0.02 uG, consistent with modeled B=B_pos at this longitude.
% Jung, McC-G et al: MHD simulations of HVCs interacting with galaxies, Magellanic system observations

%\citep{korochkinetal2025}: Can also model the Fan Region as result of large-scale structure in Local and Perseus arms, only fitting with a varying magnetic field along the cross-section of the Perseus arm.

%\citet{paveletal2012} compared NIR starlight polarization measurements at various latitudes ($-75^{\circ} < b < 10^{\circ}$ at longitude $\ell = 150^{\circ}$ with models of regular magnetic field and concluded that all (regular field only) models deviated $>3$~S/N from the models. Only Disk Even Halo Odd really ruled out, disk-even models way more probably than disk-odd. Pitch angle is $p=-6^{\circ} \pm 2^{\circ}$. Changes in magnetic field orientations are tentatively interpreted in terms of magnetic field reversals or the Galactic warp.

%%%%%%%%%%%%%%%%%%%%%%%%%%%%%%%%%%%%%%%%
\section{Small-scale magnetic fields}
\label{s:smallscale}
 In Milky Way magnetic field studies, the term ``small-scale magnetic fields'' is commonly used for magnetic fields that are not connected to any galactic large-scale features such as spiral arms or mean-field dynamo modes. 
 Small-scale magnetic fields in the Milky Way can be traced through fluctuations in Galactic synchrotron emission and/or Faraday rotation, through fluctuations in polarization angle in polarized dust emission and/or optical/NIR polarization of starlight, or through depolarization of radio synchrotron radiation, submm dust emission, and/or optical/NIR stellar polarized emission. 
 These small-scale fluctuations in the Galactic magnetic field are caused by any dynamical process in the ISM that is coupled to the field. As energy densities of gas and magnetic field are comparable, feedback mechanisms are crucial and magnetic field is an active agent in, rather than a passive tracer of, gas motions. 

Turbulent magnetic fields can arise due to the tangling of the large-scale field in interstellar turbulence or self-generation by turbulence in a fluctuation dynamo (Sect.~\ref{s:origin}). Small-scale fields are also created by compression or distortion of ambient magnetic fields by e.g.\ collapsing molecular clouds, shock waves, jets from young stellar objects, supernova remnant expansion, etc. And vice versa, gaseous and dust structures can be shaped by strong magnetic fields, creating alignments in direction between fields and filaments. Descriptions in terms of their statistical properties such as the commonly used power spectral indices, correlation scales, and sonic and magnetic mach numbers, are given in Sect.~\ref{s:ss-turb}, while Sect.~\ref{s:ss-comp} discusses the correlations with other interstellar components. This distinction into two separate sections is somewhat artificial, as magnetic field structures are expected to be affected by turbulence and interaction with discrete gaseous or dust structures simultaneously and the one can feed back to the other. It is only a way of categorizing, not implying separated, independent mechanisms in reality.

%%%%%%%%%%%%%%%%%%%%%%%%%%%%%%%%%%%%%%%%
\subsection{Magnetized turbulence}
\label{s:ss-turb}

Magnetized interstellar turbulence is exceedingly complex: it is partially compressible, non-gaussian, spatially intermittent; it is anisotropic, increasingly so towards smaller scales \citep{goldreichsridhar1995}, and exists in a multi-phase medium, which means that its properties are spatially variable, depending on environment. It can be caused by turbulent dynamo action \citep{wilkinetal2007} or shock compression \citep{bykovtoptygin1985}, often related to stellar feedback mechanisms (with the dominant contribution from supernova remnants \citep{maclow2004}), but it can also be driven by instabilities such as the magneto-rotational instability \citep{balbushawley1991} or the streaming instability \citep{kulsrudpearce1969}.
% streaming instability: CRs can self-generate alfven waves through the streaming inst.

Ideal isotropic incompressible turbulence is well described by a power spectrum, i.e.\ by the power spectral index and correlation scales. However, this treatment fails to capture the complexities of non-gaussianity such as anisotropy or intermittence known to exist in the magnetized interstellar medium. 
Various more sophisticated tools characterize various aspects of the complexity. Most work was done on neutral gas and molecular clouds due to the availability of high-resolution imaging and velocity spectral information, for example changing topology as a function of scale or density \citep[e.g.,][]{rosolowskyetal2008}, probing intermittency in molecular clouds by non-gaussian wings in velocity probability distribution functions and imaging \citep{falgaroneetal2009}, or comparisons of solenoidal and compressible modes of turbulence \citep{orkiszetal2017}. %ALSO: PCA, clumpfinding, etc
For the magneto-ionic ISM, higher-order statistical moments are used, such as higher-order structure functions \citep{setaetal2023} or bispectra which preserve phase information of the turbulence \citep{burkhartetal2009}. However, uncertainties increase with increasing orders of moments, and especially in limited data sets outliers can have disproportionally high  influence. This disadvantage is largely overcome using wavelet transforms; in particular, \citet{allysetal2019} apply the Reduced Wavelet Scattering Transform to the magnetized ISM, which describes both scales and directions with a limited number of parameters.

Observational characterization of turbulent magnetic fields is equally complicated due to the indirect nature of magnetic field measurements, the dependence of the observables on other interstellar components such as cosmic ray, gas and dust distributions, and the complexity of appropriate turbulence parameters. Observational studies typically estimate statistical turbulence parameters with the caveats that these measures are line-of-sight integrated, weighted by distributions of other species, and only probe one component of the magnetic field. 

Some characterization of small-scale magnetic fields is obtained from parametric global Galactic magnetic field models. 
Early parametric models of large-scale Galactic magnetic fields tended to include small-scale magnetic fields as gaussian random fields with a Kolmogorov \citep{kolmogorov1941} spectral index \citep{fauvetetal2011,sunetal2008}; fluctuations in Rotation Measure and polarized intensity due to these turbulent fluctuations would average out over large path lengths. These studies find widely varying ratios of turbulent-to-regular magnetic field, from negligible to dominant turbulent fields. 
A more sophisticated treatment of the turbulent magnetic field component is inclusion of an anisotropic random magnetic field component \citep{jaffeetal2010,janssonfarrar2012b}. However, these parametric models are still not able to include increasingly sophisticated descriptions of anisotropic, incompressible turbulence \citep[e.g.,][]{ferriere2020, beattieetal2025}.

Non-parametric modeling uses inversion methods to directly derive magnetic field structure as a function of distance and location on the sky, using assumptions about the nature of the structure. Various methods based on Bayesian inference have been well-developed for inversion of scalar data such as the dust distribution in the Galactic neighborhood \citep{rezaeikhetal2017,lallementetal2019,leikeensslin2019}. However, applying inversion methods to vector distributions like magnetic fields is still a huge challenge. Early efforts trace back trajectories of UHECR's through the Milky Way, using Bayesian inference to reconstruct the Galactic magnetic field in a few specific directions \citep{tsourosetal2024}. Similarly, $B_{\parallel}$ can be traced back in 3 spatial dimensions in the Solar neighborhood using forward modeling of RM based on an electron density model and the Faraday Sky \citep{mccallumetal2026}.

%Allys19:  and of the relative orientation of inter- stellar filaments and magnetic fields with dedicated histograms (Planck Collaboration Int. XXXV 2016) or more evolved tools (Jow et al. 2018).Other statistical descriptions are not specifically designed to test a phenomenological model, but rather to characterise the morphology of the observed fields1. In this category, we find descriptions in terms of filaments, sheets, and voids based on Morse theory (Sousbie 2011), or detections of linear structures, such as the Rolling Hough Transform (Clark et al. 2014).

%%%%%%%%%%%%%%%%%%%%%%%%%%%%%%%%%%%%%%%%
\subsubsection{Description in terms of power spectra}
\paragraph{Power law spectral index}
Power spectra of diffuse Galactic synchrotron emission carry information on the underlying magnetized turbulence, and are relatively easy to measure. However, the diffuse synchrotron emission also depends on the cosmic-ray electron spectrum and spatial distribution, and the orientation of any large-scale magnetic field component. Therefore, it is no surprise that measurements of the synchrotron power spectral index give variable results, depending on probed sky coverage, frequency range and multipole range (see \citealt{chatterjeeetal2025} for an overview of these measurements). \cite{lazarianpogosyan2012} introduced a theoretical formalism to describe synchrotron spectral index in terms of magnetic field fluctuations without assumptions about the spectral index of the cosmic-ray electron distribution or isotropy. They concluded that the synchrotron spectral index depends more strongly on the amplitude than on the spectral index of the cosmic ray electron spectrum, and that synchrotron fluctuations are increasingly anisotropic towards smaller spatial scales. %, and that anisotropy can be probed using higher-order statistical moments of the synchrotron structure function. 
\citet{padovanietal2021} show that the energy dependence of the cosmic-ray spectrum is essential to interpret the synchrotron spectral index, and provide a look-up table on how $B_{\perp}$ depends on the cosmic-ray electron spectral index and observing frequency, tested on MHD simulations. 

%Lazaryan \& Shutenkov (1990) used the autocorrelation function of synchrotron radiation to derive a typical scale of 90 pc at a distance of 1 kpc in a region near the north Galactic pole. They did not know distance so assumed 1 kpc. No Kolm spectrum, but spectrum is roughly consistent with $l^{3/2}$.

Similarly, power spectra of Faraday rotation measurements can be used to probe the turbulent magnetized ISM. However, RMs of extragalactic sources and pulsars are irregularly distributed over the sky, which induces significant artifacts in the Fourier analysis to obtain a power spectrum. Instead, (second order) structure functions are a more useful tool; slopes of structure functions and power spectra are linearly related. The $n$-th order structure function of RM is defined as
\[
\text{SF}(\delta\theta) = \langle \text{RM}(\theta)-\text{RM}(\theta+\delta\theta)^n\rangle_{\theta}
\]
for angular distances between sources $\delta\theta$.
Typically, the slope of these structure functions is shallower than Kolmogorov turbulence \citep{haverkornetal2006,haverkornetal2008,royetal2008,stiletal2011} and is direction dependent \citep{simonettietal1984} (although Kolmogorov-like spectral indices have also been found using RM structure functions in an H~{\sc ii} region \citep{raychevaetal2022} or polarization gradients \cite{zhangliu2025}). Some studies report broken power laws, which can be interpreted as the transition from 3D to 2D structure \citep{minterspangler1996}, discrete structures \citep{stiletal2011}, superposition of two Faraday screens \citep{haverkornetal2004_sf}, or different power laws for magnetic field and electron density \citep{xuzhang2016}. The power spectrum of the distribution of magnetic field $B_{\parallel}$ itself is also flatter than Kolmogorov ($-2.73 \pm 0.19$) in the Solar neighborhood \citep{mccallumetal2026}.
Under the assumption that neutral hydrogen filaments trace magnetic field (see Sect.~\ref{s:ss-comp}), anisotropic magnetized turbulence is probed by observed direction-dependence of measured power spectral indices of structure in neutral hydrogen filaments \citep{kalberlakerp2016,kalberlaetal2017}.
%Note that the spectral index of a RM structure function is related to the spectral index of magnetic field, but not necessarily the same (ref Lazarian??). 
The only direct measurements of interstellar turbulent magnetic field come from the Voyager satellites and reveal weak, compressible turbulence in the local interstellar medium, showing a Kolmogorov spectrum \citep{burlagaetal2015}.
%Burlaga: weak compressible turbulence with Kolm spectrum

%Roy et al. 2008: RM SFs in direction of Gal Center. Outer scale 40 pc, shallow slope
%Stil et al 2011: RM SF slopes higher close to plane, but always shallower than Kolmogorov. Interpreted as closer to plane = more local structure. Some SFs show break in spectrum - could be correlation lenhgth or concrete structures; data suggests both. 

%%"The correlation length of magnetized turbulence is observed to be around 100~pc for the Galactic halo" turns out to be hard to prove with any confidence: 
%% Lazaryan \& Shutenkov 1990: derive correlation lengths from radio internsity variations towards Noth Gal pole, but only in angular scale as the distance is unknown.
%% Ohno \& Shibata 1993: differences in RMs and DMs of pulsar pairs nearby in the sky. Assume single cell size turbulence and find 10-100 pc.
%% Chepurnov and Lazarian 2010) match WHAM electron density fluctuations to smaller scales from scattering etc and find a continuous power law uyp to ~10^17 m. No Bfield measurements
% Rand & Kulkarni 1989: residual RMs in pulsar RMs after subtraction of large-scale field: single-cell size model gives L  55 pc.

%% correlation lengths
\paragraph{Correlation length}
Next to the power spectral index, correlation length is another important parameter characterizing interstellar turbulence. 
Correlation lengths are found to be in the order of a few tens of parsecs to about a hundred pc, from dispersion in RMs of pulsars or extragalactic sources (calculated with single-cell-size models, \citet{randkulkarni1989,ohnoshibata1993,maoetal2010}). Careful treatment of turbulent magnetic fields in the \citet{janssonfarrar2012} model also shows best fits for a large correlation length of $\sim220$~pc \citep{becketal2016}. 
However, smaller RM correlation lengths are found ($\sim10-20$~pc) in the Galactic spiral arms \citep{haverkornetal2008,iacobellietal2013} and closer to the Galactic Center \citep{royetal2008,livingstonetal2021}. Note that these measurements of RM correlation lengths do not necessarily imply a similar correlation length of magnetic field, which can be significantly different \citep{hollinsetal2017}. 
The correlation length of fluctuations in magnetic field itself, obtained from modeling combined with pulsar RM, DM observations, is found to be a few tens of parsec \citep{dhakalseta2025}.

%Roy et al. 2008: RM SFs in direction of Gal Center. Outer scale 40 pc, shallow slope
% Malkovetal2010 : anisotropies in Galactic CRarrival times indicate maximum scale of 1 p, which is max scale on which particle-wave interactions are relevant -- is this the same as outer scale?
% Hollins (Shukurov) et al 2017: MHD simulations testing correlation scales, on order of tens of pc. Show that scales of RM are different fro scales of B and n_e
%previous works take ensemble averaging of fluctuations, but Beck et al (2016) show that careful consideration of small-scale turbulent fluctuations is impereative. They use an ensemble of random field realizations (instead of what was usually done using 1 realization and ensemble-averaging over many lines of sight), and calculate turbulent field effects for each realization instead of taking ensemble average of random field and including that in models. They find best-fit correlation length of 220 pc

\paragraph{Turbulent magnetic field strength}

Estimates of magnetic field strengths from the heuristic global magnetic field models give widely varying results depending on model assumptions and input data: ratios of random to regular magnetic field strength vary from $\sim0.25$ to several \citep{mivilledeschenesetal2008,jaffeetal2010,fauvetetal2011,sunetal2008}. This can likely be explained by the necessarily simplified assumptions used to describe the turbulent magnetic field, the variety in data sets, and different components included in these models. 

Detailed studies using depolarization of synchrotron radiation or variance in RMs in smaller fields result in values for turbulent magnetic field strength between one and a few microGauss \citep{gaensleretal2001,maoetal2010,schnitzeleretal2007}. 

The variety in these methods and resulting magnetic field values indicate that the small-scale magnetic field strength typically may be up to a factor of a few higher than the large-scale magnetic field strength. However, it also implies that turbulent field strength is a much too simplified diagnostic for the small-scale magnetic field across the Galaxy. More sophisticated methods of characterizing the turbulent magnetic field are captured in numerical simulations, an overview of which is beyond the scope of this observational paper.

\subsection{Correlations with other interstellar components}
\label{s:ss-comp}

One could say that if the last decade was the period of large-scale Galactic magnetic field modeling, this decade sees major and exciting progress on correlations of Galactic magnetic field structure with many, if not all, other interstellar medium components. An increasing amount of impressively detailed observations of polarization measurements combined with interstellar gas and dust components have shifted the focus from a statistical global description of interstellar magnetic fields and turbulence towards more zoomed in, detailed descriptions of correlations of magnetic field (tracers) with other components.  

\subsubsection{Correlations between magnetic field orientation and atomic hydrogen filaments}

Interstellar dust and atomic hydrogen are known to be well mixed in the interstellar medium. In addition, almost all interstellar gas has a (partial) ionization degree large enough to enable flux-freezing of magnetic field in the gas. Therefore, it may be expected that some correlation between orientations of magnetic field as probed by dust (submm dust emission or optical/NIR dust absorption) and orientations of H~{\sc i} filaments would exist.

Indeed, dust intensity ridges are aligned with magnetic field orientation from dust polarization at intermediate and high Galactic latitudes (column densities $\lesssim 10^{22}$~cm$^{-2}$), however they tend to be perpendicular to magnetic field orientation in molecular clouds \citep{planckXXXII2016,planckintXXXV2016}.

%% HI filaments and magnetic field from dust
Similarly, correlations between H~{\sc i} filaments and magnetic field orientation are ubiquitous. \citet{clarketal2014} noted that slender, long filaments in Galactic neutral hydrogen ('fibers') correlate with the magnetic field orientation as probed by interstellar polarization of starlight, which were likely originating in the Local Bubble (Sect.~\ref{s:lb}). These results were generalized in \citet{clarketal2015}, who showed correspondence between neutral hydrogen filament orientations with Planck 353~GHz dust polarization orientations at intermediate and high latitudes, concluding that it is possible to trace magnetized dust CMB foregrounds with neutral hydrogen. This alignment between H~{\sc i} filaments and magnetic field orientation was confirmed through other tracers such as Centroid Velocity Gradients 
\citep{yuenlazarian2017}, anisotropies in the distribution of spectral power correlated with magnetic field orientations \citep{kalberlakerp2016}, and polarization gradients \citep{campbelletal2022}, making these H~{\sc i} filaments a new tracer of magnetic field orientation \citep{clarkhensley2019}. 

The alignment with magnetic fields is found to be the strongest for the Cold Neutral Medium (CNM) component \citep{kalberlaetal2017,braccoetal2020}, leading to the idea that the CNM may be more magnetically aligned, and forming out of a more disordered magnetized Warm Neutral Medium (WNM) gas \citep{leiclark2024}. However, some neutral hydrogen filaments are also aligned with filaments in Faraday depth \citep{vanecketal2019}, indicating a connection between the neutral and ionized gas components as well. 

A measured correlation between the total intensity and the B-mode polarization component \citep{planckXI2020} indicates a small \citep[$\sim 2^{\circ}$,][]{cukiermanetal2023} but systematic misalignment between observed dust filaments and submm polarization. Several origins have been put forward to explain this positive TB signal: the helicity of the Galactic magnetic field \citep{braccoetal2019}, misalignment between H~{\sc i} filaments and their plane-of-sky projected magnetic field orientations \citep{clarketal2021}, the nature of the morphologies in the ISM \citep{halaletal2024a}, or line-of-sight projection effects of magnetic fields in different phases of the local ISM \citep{braccoetal2026}.

%Planck Collaboration Int. XXX 2016: EE and BB power spectra
%huffenberger et al 2020: model TB in the foreground, but vey technical 'artifically skweing distribution' so no physical explanation.

%Planck at 353 GHz (Planck Collaboration XXXII 2016) disclosed that the CNM filaments are well aligned with the thermal dust emission and with the magnetic field direction measured by Planck
%Lei and Clark 2024: CNM is more magnetically aligned,  'forming out of a disordered WNM'. "an increased magnetic field disorder in the warm neutral medium (WNM) relative to the CNM best explains the positive p 353–f CNM correlation in diffuse regions. Modeling the CNM-associated dust column as being maximally polarized, with a polarization fraction p CNM ∼ 0.2, we find that the best-fit mean polarization fraction in the WNM-associated dust column is 0.22p CNM"

%%%%%%%%%%%%%%%%%%%%%%%%%%%%%%%%%%%%%%%%
\subsubsection{Correlations with Faraday rotation and depolarization structures}

In addition to the correlations described above, many studies find correlations between Faraday rotation and radio-polarimetric structures on the one hand, and neutral hydrogen or dust filament orientations on the other, which is less straightforward to understand as Faraday rotation generally traces ionized gas. In addition, structures in Faraday rotation indicate a common strength of the magnetic field component \emph{along the line of sight}, while submm dust polarization traces the magnetic field orientation \emph{in the plane of the sky}. A correlation may be expected, but more involved and less intuitive.

A good example is the well-studied high-latitude field around the bright quasar and radio calibrator source 3C~196, which has bright and conspicuous low-frequency radio polarized structures \citep{jelicetal2015}, aligned with polarization angles from submm dust emission \citep{zaroubietal2015}. These filaments appeared to also be aligned with H~{\sc i} filaments and with low-frequency depolarization canals \citep{jelicetal2018}. Numerical simulations trying to reproduce these data hint at its use for probing magnetic field morphology in the multi-phase ISM, but also show that accurate reproduction of the observational characteristics is difficult \citep{beratetal2026}. At low radio frequencies (such as the LOFAR data used for these studies), Faraday rotation can cause total depolarization in ionized gas, so that the remaining Faraday rotation signature actually originates in neutral gas \citep{vanecketal2017}. Histograms of Oriented Gradients (HOG, \citep{soleretal2019}) analysis shows that polarized emission is indeed correlated primarily with Cold and Lukewarm Neutral Medium components, but also partially with ionized gas \citep{braccoetal2020}. These conclusions also hold for the immediate surroundings of the 3C~196 field, although dominating magnetic field orientations and polarization morphology vary \citep{braccoetal2020,turicetal2021}.

% 3C196:
%Jelic+15: LOFAR detection
%Zaroubi+15: LOFAR + dust followup
%Jelic+18: depol canals + HI also aligned
%Berat+26: hard to reproduce with num sims

% Fields around 3C196
% Bracco+20: statistical correlation between LOFAR polarised emission and HI TB. Strong correlation in Field A but nothing in field B+C. Explained by patchy morphology of polarization in B+C. "We study four fields of view - Fields 3C196, A, B, and C - and find, in at least the first two, a significant correlation between the LOFAR and EBHIS data using the histograms of oriented gradients (HOG) feature. The absence of a correlation in Fields B and C is caused by a low signal-to-noise ratio in polarization."
%Turic+2021: pol canals, compared to HI filaments and dust pol orientation. Finds correlation in fields A+B.

%vaneck+2017: IC342, and depol in ionized gas
%vaneck+2019: LoTSS diffuse pol, filament with HI and RM
%vaneck+2021: presenting new RM catalog from CGPS data

On larger scales, \citet{ercegetal2024a} conclude that this alignment is not universal over the sky --- regions where they do find alignment between neutral filaments and polarized magnetic field tracers are nearby and believed to be associated with the Local Bubble wall. However, the warm neutral medium does produce a Faraday rotation signal across that field, even if there is no visible alignment of filaments \citep{boulangeretal2024}.

\subsection{Variations in Galactic magnetic field along lines of sight}
%%%%% starlight pol and submm pol

A few of the methods mentioned in Sect.~\ref{s:methods} have the ability to distinguish magnetic field measurements along the line of sight: pulsar RMs, Faraday Tomography, and interstellar polarization of starlight. Pulsar RMs give estimates of the strength of $B_{\parallel}$ integrated over the distance to a pulsar, and observations of many pulsars in the same part of the sky therefore allows to seen trends in $B_{\parallel}$ variations along the line of sight \citep{han2017,noutsos2013,curtinetal2024}. Faraday Tomography allows distinguishing synchrotron emitting components with different Faraday depths, which carry information about $B_{\parallel}$ of which generally only the (line-of-sight integrated) field direction can be reliably determined, unless the Faraday depths can be connected to known structures. Interstellar polarization of starlight gives information on the orientation of the $B_{\perp}$ component, in subsequent dust clouds along a line of sight. With the increasing abundance of interstellar polarization of starlight measurements towards the more diffuse ISM (as opposed to focusing on dense clouds), combined with stellar distances from the Gaia mission \citep{gaiacollaboration2016}, first attempts are being made to map out variations in magnetic field orientation in dust clouds in 3D space. Lastly, dust submm polarization and line information from HI, $^{12}$CO and $^{13}$CO can be combined to estimate orientations of $B_{\perp}$ in subsequent dust clouds \citep{ferriereetal2026}.

\paragraph{Alignment of polarization angle with the Galactic plane}
%mostly along the plane
Roughly speaking, the orientation of the plane-of-sky magnetic field component as observed from interstellar polarization of starlight is largely along the Galactic plane, and increasingly so at longer distances \citep{uppaletal2024}, as one would expect from dynamo models and arguing that turbulent field structure is averaged out along sufficiently long lines of sight in the plane \citep{fosalbaetal2002}, or at higher Galactic latitudes \citep{berdyuginetal2004}. 

%but not quite along the plane
However, on smaller angular scales, magnetic field orientations are observed to significantly deviate from Galactic plane alignment in optical stellar polarimetry from the optical Interstellar Polarization Survey \citep[IPS,][]{versteegetal2023}, and NIR stellar polarimetry from the Galactic Plane Infrared Polarization Survey \citep[GPIPS,][]{clemensetal2020}, with deviations of tens of degrees on typically degree angular scales. IR stellar polarimetry towards (in front of and behind) the Galactic Center finds that most sight lines exhibit polarization angles that only slightly deviate form a plane-oriented magnetic field, although polarization angles in a few sight lines deviate by more than $45^{\circ}$ from the plane \citep{zenkoetal2020}. Locally, the Galactic magnetic field is aligned with the stellar Radcliffe wave \citet{panopoulouetal2025} and roughly aligned with the Local Bubble wall (Sect.~\ref{s:lb}).

% mostly too in dust submm pol, but not always
These deviations in magnetic field orientation with respect to the Galactic plane should correlate with submm dust polarization orientations, as both tracers probe (mostly) the same dust distribution. However, this correlation should decrease for increasingly different line-of-sight lengths in the two tracers. Indeed, \citet{planckintXIX2015} showed that the submm data in the first quadrant have an average polarization angle orientation not consistent with the infrared GPIPS results, and explain this by the longer path lengths towards Planck emission than the infrared data path lengths. On the other hand, the ratios of polarization degree of submm and infrared polarization of the farthest detectable stars suggest that path lengths \emph{are} fairly comparable in the first quadrant \citep{versteeg2025}. 

The common deviations of both optical/NIR interstellar polarization of starlight and submm dust polarized emission orientations from the Galactic plane orientation are remarkable, considering the long lines of sight involved, typically up to a few kpc. This indicates that the deviations from Galactic plane orientation must either be due to the magnetic field orientation in a dominating local feature, or must be prolonged over a significant distance along the line of sight. 

\paragraph{Galactic magnetic field tomography based on interstellar polarization of starlight}
% in clouds
Interstellar polarization of starlight provides a way to probe the orientation of the plane-of-sky magnetic field component \emph{locally} in dust clouds along the line of sight, if one can disentangle various (foreground) components along the line of sight. Provided that the magnetic field orientation changes with each dust cloud, the polarization properties of stars in front of and behind this cloud will change. Therefore, observations of observed polarization changes at a certain distance indicate that there is a dust cloud at that distance. 

Using this optical starlight tomography method, care has to be taken to subtract polarization signal from foreground clouds. Several methods have been used for this foreground subtraction, such as $\chi^2$-like minimization \citep{anderssonpotter2006}, maximizing the signal-to-noise ratio of polarization degree of a far cloud as a function of assumed cloud distance \citep{panopoulouetal2019}, breakpoint analysis detecting abrupt changes in polarization as a function of distance \citep{doietal2021}, clustering algorithms \citep{versteegetal2024}, bayesian inference \citep{pelgrimsetal2023}, or changes in the slope of cumulative Mahalanobis distances\footnote{The Mahalonobis distance is used here as a measure of the probability distribution of the combined Stokes $Q$ and $U$ distributions.} of Stokes $Q$ and $U$ \citep{uppaletal2026}. \citet{angaritaetal2025} showed that clustering algorithms and bayesian inference gave consistent results. Note that these methods rely on variations in polarization along the line of sight: multiple clouds along the line of sight with constant magnetic field orientation, and hence constant polarization, are undetectable through their polarization properties alone.

In all of these studies, large deviations of polarization angles from the Galactic plane orientation and from cloud to cloud are ubiquitous \citep{panopoulouetal2019,doietal2024,versteegetal2024,pelgrimsetal2023,angaritaetal2025}. Polarization angles are mostly \citep[but not always, see][]{angaritaetal2025,pelgrimsetal2023} coherent in the plane of the sky, which can be explained by the small size of the fields which typically probe a fraction of a parsec. In the only study that analyzes a larger field of view of 4 square degrees, coherence in polarization angles is seen up to a scale of $\sim20$~pc \citep{pelgrimsetal2023}. %3 degrees at a distance of 374 pc
The emerging image of variable polarization angle orientations in multiple dust components along the line of sight was also noted by \citet{halaletal2024b}, who find that the polarization fraction of Planck dust emission data decreases for an increasing number of dust components along the line of sight.
All results seem qualitatively consistent with a turbulent magnetized ISM; however, in those clouds associated with the Local Bubble, more coherent structure is also observed (see Sect.~\ref{s:lb}).

\section{The Local Bubble}
\label{s:lb}
% Hot gas still debated?
%Galeazzi et al (2014) say yes hot gas, but see Clark et al (2014).
% Hot:  (Puspitarini et al. 2014; Galeazzi et al. 2014; Snowden et al. 2015; Liu et al. 2017)

%Erceg+2024a: tracing differences in dust, HI filaments, starlight optical linear polarisation, and depolarisation canals. No evidence for universal alignment. Found alignemtn in one region, starlight pol says 200-240 pc, so edge of LB
%Erceg+2024b: new mosaic from LoTSS-DR2. Ordered/aligned structure over 2/3 of field. Probably LB wall but maybe also from medium inside LB (clouds). Cannot model with neutral gas only, need an ionized front.

The Local Bubble, also commonly called Local Cavity or Local Chimney, is a low-density cavity with the Sun roughly in its center, as first discovered due to its low extinction \citep{fitzgerald1968}. It is thought to originate from clustered supernova explosions about 14~Myrs ago \citep{zuckeretal2022}. The bubble is mostly filled with hot, dilute gas, although small amounts of denser neutral or ionized gas (clouds) and dust manage to survive in the hot environment \citep{frisch2007,farhangetal2019,zuckeretal2025}. Matter and magnetic field are compressed into a denser shell of irregular shape, including a ring of star forming regions \citep{zuckeretal2022}.
Inversion methods using dust extinction or color excess show a shell with varying distances to the Sun, from $\sim50$~pc to $\sim150$~pc \citep{vergelyetal2010,lallementetal2019}, or even higher in particular directions \citep{oneilletal2024}.
%The dust structure \citet{edenhoferetal2024} is mostly but not completely consistent with neutral gas maps as traced by the Na~{\sc I} doublet \citep{lallementetal2003,welshetal2010}, which tend to show a slightly smaller bubble. Partially ionized gas from Ca~{\sc II} observations is detected coincident but also in front of the neutral gas and dust structures \citep{welshetal2010}.

%Figure?
These tracers show that the Local Bubble is elongated in the direction roughly perpendicular to the Galactic plane, and is shaped very irregularly due to interaction with gaseous structures such as the walls of neighboring bubbles, "interstellar tunnels" connecting to these bubbles, and chimneys in which hot gas escapes from the Galactic disk \citep{lallementetal2003,oneilletal2024}. The embedding and interaction of the Local Bubble wall with other structures can make it difficult to distinguish it properly, especially at lower Galactic latitudes \citep{gontcharovmosenkov2019,versteegetal2024}.

In principle, the magnetic field in the Local Bubble wall is a very local feature and not directly part of the general magnetic field in the Milky Way. However, since all observations of magnetic field tracers probe through the Local Bubble wall, its magnetic field influences \emph{all} data used for investigation of Galactic magnetism. The Local Bubble wall is the dominant source of magnetically-aligned neutral hydrogen fibers \citep{clarketal2014}, and is a significant contributor to polarized dust emission \citep{skalidispelgrims2019} and Faraday rotation \citep{reissletal2023,ercegetal2024b} at intermediate and high latitudes. Including the Local Bubble magnetic field in large-scale heuristic models pre-empts the need for an anisotropic random field component in the Galactic magnetic field \citep{korochkinetal2025}. Therefore, a discussion of current knowledge of the magnetic field of the Local Bubble wall is warranted here.

The larger-scale magnetic field in the Solar neighborhood is oriented along the Local Arm (also called Local Spur, Orion Arm) at Galactic longitude $\ell\sim 80\dg-90\dg$ \citep{randlyne1994,heiles1996,xuhan2019,hutschenreuterensslin2020}, consistent with galaxy-scale models of magnetic fields following spiral arms \citep{jaffe2019}. However, the magnetic field in the Local Bubble wall does not follow this large-scale field but is deformed \citep{leroy1999}, roughly consistent with a deformation due to a bubble blown in a uniform magnetic field \citep{heiles1998}, as predicted by models of expanding supernova-blown bubbles \citep{stiletal2009}. 

%Planck (Planck intermediate results XLIV. Structure of the Galactic magnetic ﬁeld from dust polarization maps of the southern Galactic cap) estimates the direction of the GMF from Planck polarization towards the Southern Galactic cap as $(\ell,b) = (70^{\circ}\pm 5^{\circ}, 24^{\circ}\pm 5^{\circ})$. 
%Frisch 2007: "Heiles et al. (1980) found a volume averaged field strength of BIS∼ 4 μG in a tangential direction through the shell (extending ∼ 70 ± 30 pc towards ℓ= 34◦, b= 42◦)": paper on NPS ISM adn magnetic field; this region is location of highest RM.

A first analytical model of the structure and strength of the Local Bubble magnetic field used an ellipsoidal model with a bubble-blown magnetic field aligned with the bubble wall \citep{alvesetal2018}. Their results are roughly consistent with Planck 353~GHz dust polarization measurements towards the Galactic pole regions ($|b| > 60\dg$), but they concluded that more complex and irregular models were needed. \citet{pelgrimsetal2020} modeled the Local Bubble shape from spherical harmonics fit to extinction measurements, combined with an analytical magnetic field model adapted from \citet{alvesetal2018}. Their derived orientation of the original large-scale magnetic field is consistent with earlier estimates. LOFAR Faraday tomography observations are consistent with a magnetic field orientation along the wall of this bubble, but need an ionized component to the bubble wall density to explain the data \citep{ercegetal2024b}. In addition, these authors find magnetic field orientations towards the Galactic polar regions with a small component parallel to the line of sight, consistent with Faraday rotation measurements. This line of modeling was continued by \citet{oneilletal2024}, who employed a new, more detailed 3D dust distribution and derived the wall's magnetic field from dust polarization measurements, assuming it is tangent to the wall's surface. They confirm a small magnetic field component parallel to the line of sight towards the galactic caps, and also find a consistent orientation of the original large-scale magnetic field.

There are multiple lines of evidence suggesting that the Local Bubble wall is a dominant contributor to polarized dust emission at intermediate and high latitudes. The ratio of dust polarization fraction over the degree of interstellar polarization of starlight flattens out behind the Local Bubble wall at $|b| \gtrsim 60^{\circ}$, suggesting that the Local Bubble wall dominates dust submm observations \citep{skalidispelgrims2019}. 
This conclusion is confirmed by the plateauing of the fraction of interstellar polarization of starlight after 150--250~pc across the
sky \citep{gontcharovmosenkov2019}. MHD simulations of a Local Bubble analogue also find that the dust polarization degree and morphology of the field lines are in qualitative agreement with their Local Bubble model \citet{maconietal2023}.

However, no correlation between observed Planck polarization degree and magnetic field inclination angle was found from comparisons of the observed Planck polarization degree and magnetic field inclination angle with respect to the Local Bubble wall in both the \citet{pelgrimsetal2020} and \citet{oneilletal2024} models \citep{halaletal2024b}. These authors conclude that either the Local Bubble cannot be the dominant contributor to the dust polarization at intermediate/high latitudes, or the assumption that the magnetic field runs parallel to the Local Bubble wall is incorrect. 

%Halal selected LOS in quartile with maximum ratio of AV_lB (integrtead Lallement from LB inner wall to 50pc ahead) to AV_planck, so where LB dominates (well, ratio > 0.2) (latutude cut >60deg does not make a difference).

%Halal et al 204: "The observed dust polarization fraction also depends on other factors, such as the dust grain alignment efficiency (King et al. 2019; Medan & Andersson 2019), the phase distribution of the neutral interstellar medium (Lei & Clark 2023), and measurement noise. However, the 3D structure of the magnetic field is one of the major factors (Clark 2018; Hensley et al. 2019; Planck Collaboration et al. 2020a)."

The magnetic field strength in the Local Bubble wall has been estimated from interstellar polarization of starlight measurements and DCF analysis as variable between $\sim 8 - 40~\mu$G \citep{anderssonpotter2006, medanandersson2019}. %two groups of Bstrengths around 8 and around 40 uG, Medan says due to compression of the LB wall.

Concluding: the dusty, magnetized wall of the Local Bubble has a significant influence on interstellar polarization measurements over a large part of the sky. Towards the galactic polar regions ($|b| \gtrsim 60\dg$), the Local Bubble wall even dominates measurements of interstellar magnetic fields through optical interstellar polarization of starlight and Faraday rotation, although some polarizing dust is coming from longer distances. The orientation of the magnetic field is roughly consistent with a field tangent to the intricately shaped Local Bubble wall, assuming that the original magnetic field in which the Local Bubble expanded has an orientation along the local Orion arm. However, there is small-scale structure in the wall's magnetic field and/or in the orientation of the wall itself.

\section{Current surveys and future prospects}
\label{s:future}

In all main tracers of the Galactic magnetic field, exciting current and near-future developments will greatly increase available data in many dimensions. Below is an attempted overview of the main projects, divided up into radio-polarimetric (RM Grids and/or Faraday Tomography of diffuse emission) and optical/NIR stellar polarization surveys that promise to shed more light on magnetism in the Milky Way in the near future.

%%%%%%%%%%%%%%%%%%%%%%%%%%%%%%%%%%%%%%%%
\subsection{RM grid surveys}

For more than a decade, the NVSS RM catalog of \citet{tayloretal2009} was the main source of extragalactic point source RMs across the global sky (although in the plane, Galactic Plane Surveys with higher RM source density existed \citep{brownetal2003,stiletal2006,brownetal2007}). With its 37,543 sources distributed over the sky, this means a RM source density of less than 1 source per square degree; using only two frequencies, RMs were vulnerable to 2-pi-ambiguities; typical errors are 1--2~rad~m$^{-2}$. 

A high source density is not only essential for probing external galaxies, galaxy clusters or the cosmic web, but also the Faraday rotation of the Milky Way shows small-scale large variations in RM \citep[e.g.][, in Sect.~\ref{s:sgrarm}]{shanahanetal2019}. In addition, high accuracy of point source RM measurements is crucial for detecting weaker structure in magnetic field and/or electron density. Several surveys are currently being executed that will significantly improve our global RM Grid on these points.

\paragraph{POSSUM}
The Polarisation Sky Survey of the Universe's Magnetism (POSSUM) is a multi-year survey with the Australian SKA Pathfinder (ASKAP) currently in its third year. It has a frequency range of primarily 800--1088~MHz, an angular resolution of $20^{\prime\prime}$ and sensitivity of $18~\mu$Jy~beam$^{-1}$.
It will produce an RM grid of the entire Southern sky with the high density of $\sim 30- 50$ RMs per square degree with a median uncertainty of $\sim 1$~rad~m$^{-2}$. Results from POSSUM pilot observations at two broad frequency bands (800--1087~MHz and 1316--1439~MHz) achieve RM source densities of $\sim 35, 30$ sources with RM uncertainty of $\sim 1.5, 13$~rad~m$^{-2}$ for the lower and upper frequency band, respectively \citep{vanderwoudeetal2024}.

\paragraph{SPICE-RACS}
The Spectral and Polarisation in Cutouts of Extragalactic sources from RACS (SPICE-RACS) project has the aim of providing an RM Grid from spectro-polarimetric Rapid ASKAP Continuum Survey (RACS) data across the Southern sky, at $15^{\prime\prime}$ angular resolution, at $\sim 200\mu$Jy/PSF rms noise, and in a 800--1088~MHz frequency range. The second data release covers the entire southern sky up to a declination of $+49^{\circ}$. This survey presents $3.4\times10^5$ RM sources above $6\sigma$, reaching a source density of $\sim 7$~RMs per square degree \citep{thomsonetal2026}.

\paragraph{VLASS}

The Very Large Array Sky Survey (VLASS) images the entire sky visible to VLA ($\delta > -40^{\circ}$) in the broad band of 2--4~GHz with a high resolution of 2.5$^{\prime\prime}$ \citep{lacyetal2020}. Data taking for the first phase (three epochs) is finished, while a fourth epoch is ongoing. About $10^5$ detected RM~soures are projected. The advantage of the higher frequency range is possible detection of sources that are depolarized at lower frequencies, and sensitivity to high RM values.

\paragraph{MMGPS}

The MPIfR-MeerKAT Galactic Plane Survey (MMGPS) is a commensal survey with MeerKAT at a frequency range of 856--1712~MHz and covers various sky areas in the Galactic plane \citep{padmanabhetal2023}. More than 5000 RM sources are expected, leading to a high source density of about 25 per square degree over selected areas of the Galactic plane. The wide frequency range of $800-3500$~GHz ensures sensitivity to Faraday-thick structures and extreme RMs.

\paragraph{LoTSS}
The LOFAR Two-metre Sky Survey (LoTSS) is a low-frequency (120--168~MHz) radio-polarimetric survey of most ($\sim 88\%$) of the Northern Sky at an angular resolution of $6^{\prime\prime}$ down to a median sensitivity of $92~\mu$Jy~beam$^{-1}$ \citep{shimwelletal2026}. An RM Grid has been constructed of the LoTSS Data Release 2, which covers 7520~square~degrees on the Northern sky, yielding 2431 RM~sources \citep{osullivanetal2023}. Due to increased depolarization at low frequencies, the source density is not the highest, but the unique strength of low-frequency RM grids is the high RM precision, with a median RM uncertainty of 0.06~rad~m$^{-2}$, not including a possible systematic uncertainty of $\lesssim 0.3$~rad~m$^{-2}$ due to the ionospheric correction.

\paragraph{MWA-POGS}

The POlarised GaLactic and Extragalactic All-Sky MWA Survey (the POlarised GLEAM Survey, POGS) has a similar low frequency range of 169--231~MHz as LoTSS, but covers almost the entire southern sky at delcination $-82^{\circ}$ to $+30^{\circ}$ \citep{riseleyetal2020}. The Phase I Faraday Rotation Measure survey has an angular resolution between 3$^{\prime}$ and 7$^{\prime}$ and contains 517 RM sources with a median RM uncertainty of 0.38~rad~m$^{-2}$.

\paragraph{SKA}
For the Square Kilometre Array (SKA), under construction at the moment, an RM Grid survey of the entire Southern sky is planned with SKA-Mid, which is expected to produce RM measurements from 2--3 million extragalactic sources, a factor of a few more than POSSUM \citep{healdetal2020}. A low-frequency RM Grid survey using SKA-Low will complement this, contributing less sources but a 100~times higher precision in RM.

%%%%%%%%%%%%%%%%%%%%%%%%%%%%%%%%%%%%%%%%
\subsection{Spectro-polarimetric surveys of diffuse emission}

For imaging diffuse polarized emission, sensitivity to large-angular-scale structure is crucial. Hence, most surveys of diffuse polarized emission are single-dish surveys, possibly with matching interferometer data to increase angular resolution (POSSUM, SKA-MID). Only at the lowest frequencies, even small RM fluctuations cause randomized $Q, U$ fluctuations on sufficiently small scales that there is no large-scale component in polarized emission left. In that case, interferometric surveys (LOFAR, MWA, SKA-LOW) can cover the complete signal. 

\paragraph{GMIMS}
The Global Magneto-ionic Medium Survey (GMIMS) endeavors to complete a spectro-polarimetric survey of the entire sky from $\sim300$~MHz to $\sim1800$~MHz. GMIMS consists of various sub-surveys in different frequency ranges and hemispheres, some of which are published, others are ongoing. Due to the differences in frequency ranges and angular resolutions, resulting in varying amounts of depolarization, the sub-surveys largely probe different sightlines through the interstellar medium \citep{hill2018}. GMIMS consists of:
\begin{itemize}
\item GMIMS-Low Band South: 300--480~MHz with Murriyang, the Parkes radio telescope \citep{wollebenetal2019};
\item GMIMS-Low Band North: 400--800~MHz with CHIME, in progress;
\item DRAGONS: 350--1030~MHz with the DRAO 15m Telescope \citep{ordogetal2026};
\item GMIMS-Mid Band South (PEGASUS, PI Carretti): 704--1440~MHz, with Murriyang, in progress;
\item GMIMS-Mid Band North: 900--1700~MHz, with the DRAO 15-m Telescope and/or John A.\ Galt Telescope, in planning phase;
\item GMIMS-High Band South (STAPS): 1328--1768~MHz, with Murriyang \citep{sunetal2025,raychevaetal2025};
\item GMIMS-High Band North: 1280--1750~MHz, John A. Galt Telescope \citep{wollebenetal2021,dickeyetal2019}.
\end{itemize}

\paragraph{S-PASS} 
The S-Band Polarization All Sky Survey (S-PASS) with Murriyang complements GMIMS at the high-frequency end for the Southern sky, in the bands of 2176--2216~MHz and 2272---2400~MHz, at an angular resolution of $8.9^{\prime}$ and a pixel rm sensitivity of 0.815~mK \citep{carrettietal2019}. It does not report Faraday spectra because the small frequency coverage would introduce significant artifacts, but calculated RM from a linear relation between angle and wavelength squared.

%\paragraph{C-BASS?} Too high: 5 GHz, almost no Faraday rotation

\paragraph{LoTSS}

Besides an RM Grid, the LoTSS survey provides spectro-polarimetry of diffuse emission in the entire Northern sky. Significant parts of the Northern sky are imaged \citep{ercegetal2022,ercegetal2024a,ercegetal2024b} at an angular resolution between $4^{\prime}$ and $5.5^{\prime}$, a Faraday depth resolution of $1.8$~rad~m$^{-2}$ and a mean sensitivity of $117~\mu$Jy~PSF$^{-1}$~RMSF$^{-1}$. Due to the high depolarization at low frequencies, these measurements probe only the very nearby interstellar medium, such as the large synchrotron loops and the Local Bubble wall.

%%%%%%%%%%%%%%%%%%%%%%%%%%%%%%%%%%%%%%%%
\subsection{Optical/NIR stellar polarization surveys}

Polarization measurements of stars have been recorded for many decades, recently assembled into an all-sky catalog of 55,742 polarization measurements from 42,482 stars, including their distances \citep{panopoulouetal2025}. This catalog, containing slightly more than one source per square degree on average, will be complemented by two surveys measuring polarization of stars at much higher source densities, PASIPHAE and SouthPol, allowing both small-angular-scale magnetic field orientations to be investigated, as well as magnetic structure resolved along the line of sight.

NIR polarized stellar measurements are advantageous as stars at larger distances can be probed; however, as both the brightness and polarization degree are lower in the NIR than in optical, observing polarized stars in NIR over large parts of the sky is difficult. The largest existing survey is the Galactic Plane Infrared Polarization Survey (GPIPS) with the 1.83m Perkins telescope \citep{clemensetal2020}. GPIPS surveyed part of the Northern Galactic plane ($18^{\circ}<\ell<65^{\circ}$, $|b|<1^{\circ}$) at $1.6~\mu$m (H-band), and cataloged more than a million stars with $m_H < 12.5$~mag and uncertainties in polarization percentage $< 2\%$.

\paragraph{SouthPol}
SouthPol aims to survey the entire Southern sky in V-band (551~nm) down to a depth of $V \approx 14-15$ with a new 1m telescope at Observat\'orio do Picos dos Dias in Brazil, resulting in millions of stars across the Southern sky \citep{magalhaesetal2005}. The pilot Interstellar Polarization Survey \citep[IPS,][]{versteegetal2023}) demonstrated that a source density of over 5000 stars per square degrees at polarization degree $\gtrsim 0.8\%$ is possible for comparable exposure times, making SouthPol an excellent resource for magnetic field tomography in the Galactic plane, probing dust clouds and star forming regions. 

\paragraph{PASIPHAE}
The Polar-Areas Stellar Imaging in Polarization High-Accuracy Experiment (PASIPHAE) plans to survey the Northern and Southern Galactic polar caps in R-band ($658$~nm), together covering $> 10,000$~square degrees, at the South African Astronomical Observatory in Sutherland, South Africa, and the Skinakas Observatory in Crete, Greece \citep{tassisetal2018}. PASIPHAE is predicted to result in over 360 measurements of stellar optical polarization per square degree with polarization percentages $\gtrsim 0.5\%$ (at $>3\sigma$). The high high latitudes are ideal for probing Cosmic Microwave Background Radiation foregrounds, but also allow investigation of Galactic magnetism out of the plane. 

%CRAFTS (Commensal Radio Astronomy FAST Survey) talks only about pulsars, FRBs, HI external galaxies, no pol ISM (Li+ 2018).

%%%%%%%%%%%%%%%%%%%%%%%%%%%%%%%%%%%%%%%%
\section{Conclusions}

Observational studies of the Milky Way's magnetic field have yielded a wealth of new knowledge --- along with new questions --- driven by major recent advances in instrumental capabilities and thus high-resolution, high-sensitivity data across large portions of the sky. Equally impressive progress was achieved through major advances in numerical simulations and theoretical studies, although these fall outside the scope of this review. Over the past decade, RM Grids have increased substantially in source density and RM precision, as have the methods used to distinguish Galactic foreground from intrinsic and other background RM components. The application of Faraday Tomography has matured from exploratory to a deeper and more nuanced understanding of its results, including its limitations and pitfalls. The Planck satellite has revealed an entirely new view on polarized dust, uncovering detailed magnetic field structures both near the Galactic plane and at high latitudes. Meanwhile, a resurgence in interstellar starlight polarization observations --- now capable of probing larger regions of the sky and detecting weaker polarization signals --- holds great promise for mapping magnetic field structure across three spatial dimensions.
 
Heuristic models of the large-scale Galactic magnetic field have confirmed the spiral structure of the field in the disk, placed significant constraints on its strength and direction, and revealed distinct magnetic field configurations in the Galactic disk and halo. However, these models are gradually approaching their limits as tools for global Galactic magnetic field modeling. The growing volume and quality of data reveals increasing complexity and deviations from large-scale models, creating the need for an excessive number of free parameters; however, nput from dynamo models may help to partially constrain these parameters.  Non-parametric models of the Galactic magnetic field circumvent these limitations, but remain in their early stages, having so far been applied only to limited fields of view and/or the near Solar neighborhood, including artifacts and/or returning only one magnetic field component.

Driven by these developments, research is increasingly focused on sophisticated methods for characterizing an anisotropic and intermittent turbulent magnetic field, as well as on the correlations between magnetic field structure and other ISM tracers. These investigations are revealing complex relationships between magnetic field structures and the multi-phase interstellar gas and dust. Both gas and dust are filamentary across a wide range of scales, shaped by the structure and orientation of the magnetic field; therefore, these filamentary structures are increasingly used as proxies for magnetic field orientation.

Numerous exciting observational developments are currently underway, generating new RM Grids, Faraday Tomography data cubes, and catalogs of optical interstellar starlight polarization across large areas of the sky. These efforts inspire well-founded confidence in equally exciting advances in the near future.

\backmatter

%\bmhead{Supplementary information}

%If your article has accompanying supplementary file/s please state so here. 

%Authors reporting data from electrophoretic gels and blots should supply the full unprocessed scans for key as part of their Supplementary information. This may be requested by the editorial team/s if it is missing.

%Please refer to Journal-level guidance for any specific requirements.

\bmhead{Acknowledgements}

The author acknowledges insightful discussions with Andrea Bracco, % (dust LB)
Susan Clark,  JinLin Han,  Yue Hu, Alex Lazarian, % (paper Halal)
%Isabelle Grenier, %(magnetic field direction in local bubble - in review?); thanks to 
Naomi McClure Griffiths, Theo O'Neill, Rafael Skalidis, and Mehrnoosh Tahani. Thank you to Anvar Shukurov for comments on parts of the paper, and likely to a number of others once this document is on astro-ph.

%Gilles Joncas: dust and gas are superwell imxed known since time of IRAS. Never seen any bubbles with separation: but I don't see separation in all parts, just in some. Na has a lower ion potential than HI, so if NaI survives, there should certainly be HI associated.

The author acknowledges the Interstellar Institute's program ``II7'' and the Paris-Saclay University's Institut Pascal for hosting discussions that nourished the development of the ideas behind this work. The author also gratefully acknowledges the hospitality of the NWO-I Institute HFML-FELIX while writing parts of this review.

\section*{Declarations}
The author acknowledges funding from the European Research Council (ERC) under the
European Union's Horizon 2020 research and innovation programme (grant
agreement No 772663). 
%Some journals require declarations to be submitted in a standardised format. Please check the Instructions for Authors of the journal to which you are submitting to see if you need to complete this section. If yes, your manuscript must contain the following sections under the heading `Declarations':

%\begin{itemize}
%\item Conflict of interest/Competing interests (check journal-specific guidelines for which heading to use)
%\item Ethics approval and consent to participate
%\item Consent for publication
%\item Data availability 
%\end{itemize}

%\noindent
%If any of the sections are not relevant to your manuscript, please include the heading and write `Not applicable' for that section. 

\section*{Declarations}

\bmhead{Conflict of interest} The author declares no conflict of interest.

\phantomsection
\addcontentsline{toc}{section}{References}
\bibliography{haverkorn-bibliography}% common bib file

\end{document}